\documentclass[a4paper,11pt]{article}
\pdfoutput=1

\usepackage{jheppub}
\makeatletter\def\@fpheader{\\}\makeatother

\usepackage{amsmath}
\usepackage{amsfonts}
\usepackage{amssymb}
\usepackage{amsthm}
\usepackage{mathtools}
\usepackage{mathdots}
\usepackage{multirow}
\usepackage{dsfont}
\usepackage{graphicx}
\usepackage{tabularx}

\usepackage{tikz}
\usetikzlibrary{arrows}

\newcommand{\op}[1]{\operatorname{#1}}

\newcommand{\ii}{\operatorname{i}}
\newcommand{\ee}{\operatorname{e}}
\newcommand{\dd}{\mathrm{d}}
\newcommand{\ppi}{\mathrm{\pi}}
\newcommand{\II}{\mathds{1}}

\newcommand{\tr}{\mathsf{{\scriptscriptstyle T}}}

\newcommand{\eJ}{\mathrm{{\scriptscriptstyle J}}}

\newcommand{\Tr}{\operatorname{Tr}}
\newcommand{\diag}{\operatorname{diag}}
\newcommand{\sgn}{\operatorname{sgn}}
\renewcommand{\Re}{\operatorname{Re}}
\renewcommand{\Im}{\operatorname{Im}}

\newcommand{\gmn}{g_{\mu\nu}}
\newcommand{\fmn}{f_{\mu\nu}}
\newcommand{\p}{\partial}

\newcommand{\mcC}[1]{\operatorname{\mathcal{C}}(#1)}

\theoremstyle{plain}
\newtheorem*{thm*}{\protect\theoremname}
\providecommand{\theoremname}{Theorem}

\newenvironment{pmatrixc}{
  \bgroup\renewcommand{\arraystretch}{1}\begin{pmatrix}
}{
  \end{pmatrix}\egroup
}

\newenvironment{pmatrixr}{
  \bgroup\renewcommand{\arraystretch}{1}\begin{pmatrix*}[r]
}{
  \end{pmatrix*}\egroup
}

\newcommand*{\doi}[1]{\href{http://dx.doi.org/#1}{#1}}

\title{On the local structure of spacetime in ghost-free bimetric
  theory and massive gravity}

\author{S. F. Hassan,}
\author{Mikica Kocic}

\affiliation{
  Department of Physics \& The Oskar Klein Centre,\\
  Stockholm University, AlbaNova University Centre,
  SE-106 91 Stockholm
}

\emailAdd{fawad@fysik.su.se}
\emailAdd{mikica.kocic@fysik.su.se}

\abstract{The ghost-free bimetric theory describes interactions of
  gravity with another spin-2 field in terms of two Lorentzian
  metrics. However, if the two metrics do not admit compatible notions
  of space and time, the formulation of the initial value problem
  becomes problematic.  Furthermore, the interaction potential is
  given in terms of the square root of a matrix which is in general
  nonunique and possibly nonreal.  In this paper we prove that the
  reality of the square root matrix leads to a classification of the
  allowed metrics in terms of the intersections of their null
  cones. Then, the requirement of general covariance further
  constrains the allowed metrics to admit compatible notions of space
  and time.  It also leads to a unique definition of the square root
  matrix. The restrictions are compatible with the equations of
  motion. These results ensure that the ghost-free bimetric theory can
  be defined unambiguously and that the two metrics always admit
  compatible 3+1 decompositions, at least locally. In particular,
  these considerations rule out certain solutions of massive gravity
  with locally Closed Causal Curves, which have been used to argue
  that the theory is acausal.}

\keywords{Modified gravity, Interacting spin-2 fields, Bimetric theory,
  Massive gravity}

\begin{document}

\maketitle
\flushbottom


%
\section{Introduction and summary}

Despite the overwhelming observational and theoretical evidence for
physics beyond the standard model and general relativity, the absence
of detailed observational data has so far hindered finding the
underlying theory through the phenomenological model building. An
alternative approach has been to investigate possible theoretical
frameworks for the new physics guided by consistency requirements such
as causal evolution and the absence of ghost instabilities, a well
developed example being string theory.

To recapitulate the situation from the low energy point of view, the
standard model contains multiplets of spin-0, spin-$\tfrac{1}{2}$, and
spin-1 fields, and the multiplet structure (with the associated
symmetries, broken or exact) are critical for the consistency and
viability of the model. In contrast, general relativity is a theory of
a \textit{single massless} spin-2 field. The low energy limits of
superstring theories and other grand unification schemes lead to
similar field contents; characterized by a single massless spin-2
field, and a proliferation of lower spin multiplets. Hence, in the
search for field theories beyond GR, a complementary approach would be
to consider theories of gravity with extra spin-2 fields, in part
motivated by the observation that many of the new physics signals are
gravity related.  Irrespective of their phenomenological relevance, it
is an interesting theoretical question to find out if such theories
could exist at all, or if GR is the unique (classically) consistent
theory of a spin-2 field.  This question has attracted considerable
attention since the early works on massive \cite{FP,BD} and
interacting spin-2 fields \cite{Isham:1971gm}\footnote{The related
  work in \cite{Zumino} had remained completely unnoticed.}, but
remains unresolved so far.\footnote{Regarding multiple massless
  spin-2 fields, it has been shown in \cite{Boulanger:2000rq} that
  such theories with non-Abelian gauge symmetries cannot exist.}  The
stumbling blocks to be avoided are instabilities, such as the
Boulware-Deser (BD) ghost \cite{BD}, and problems with causality due
to nonminimal couplings of the metric.  Ghost-free theories of massive
gravity \cite{dRGT,HR1106} and interacting spin-2 fields
\cite{HR1109,HR1111} have now been constructed, but the complete
classical consistency of these theories needs to be further
investigated.

In this paper we consider the ghost-free bimetric theory \cite{HR1109}
which is a theory of the gravitational metric $\gmn$, in the presence
of an extra spin-2 field $\fmn$, and without the BD ghost. We address
two potentially debilitating problems that are inherent in the
construction of such theories and show that both are avoided by the
natural requirements of the reality of the action and its general
covariance. Our results also apply to ghost-free massive gravity
\cite{dRGT,HR1106} which is obtainable from bimetric theory on
freezing the dynamics of $\fmn$. However, in this case the
implications are somewhat weaker. Here we briefly elaborate on
the problems to be resolved and then present a brief summary of our 
results.

In formulating the ghost-free bimetric theory one encounters two
potential problems that could render the theory ill-defined from the
outset. One of the problems relates to the compatibility of the
notions of space and time for the two metrics $\gmn$ and $\fmn$. Since
each metric has an Einstein-Hilbert term in the action, as dictated by
the absence of ghost, they come with their respective
notions of space and time which, a priori, may not be compatible with
each other. For example, if in any 3+1 decomposition a common
spacelike hypersurface does not exist, then formulating the dynamics
of the theory as time evolution of initial data on a spatial
hypersurface becomes problematic from the outset. 

Another problem is that the interaction potential between the metrics
$\gmn$ and $\fmn$ is given in terms of a matrix $S^\mu{}_\nu$ defined
through the equation $S^2=g^{-1}f$. The solutions are the matrix
square roots $S=\sqrt{g^{-1}f}$, which are nonunique and possibly
nonreal. To define the theory unambiguously, one must require $S$ to
be real and also provide a criterion to specify it uniquely. Then one
has to find the corresponding restrictions on $\gmn$ and $\fmn$ and
ensure that they are consistent with the dynamics.

\subsection{Summary of results} 
The main result of this paper is a theorem showing that the reality
condition on the equations leads to a classification of the allowed
$\gmn$ and $\fmn$ configurations based on how their null cones
intersect. Then compatibility with general covariance further
restricts the allowed configurations to metrics $\gmn$ and $\fmn$ that
admit compatible notions of space and time, consistent with the
equations of motion. General covariance also leads to a unique
definition of the square root matrix. This is elaborated below.

To ensure that the bimetric equations
always yield real $\gmn$ and $\fmn$ as solutions, we require that only
configurations for which the matrix $g^{-1}f$ admits real square roots
are allowed. This requirement is converted into conditions on $g$ and
$f$, and leads to four classes allowed metric configurations
classified as Type I to Type IV (including some subtypes).  Each
metric Type has a geometric significance. In Type I-III
configurations, the metrics $\gmn$ and $\fmn$ have null cones that
always intersect such that they have common timelike and common
spacelike directions, hence the metrics admit compatible notions of
space and time. In Type IV, the null cones touch only along two null
directions (the simplest example being $g=\diag(-1,1,1,1)$,
$f=\diag(1,-1,1,1)$) and the metrics do not admit compatible notions
of space and time. In all other cases where the null cones do not
intersect as above, $g^{-1}f$ does not admit any real square root.

For Type I-IV configurations the real square roots $S=\sqrt{g^{-1}f}$
are not unique. In particular, Type IV metrics only admit (an infinite
number of) \textit{nonprimary} real square roots. For nonprimary roots,
$S^\mu{}_\nu$ does not transform as a (1,1) tensor and the theory
will violate general covariance. To show that other configurations
cannot dynamically evolve into Type IV, we note that, for Type IV
metrics, $S$ coincides with branch cuts of the square root function
and, around such configurations, the variation $\delta S$ is
ill-defined. Hence such metrics are also excluded by the variational
principle and the correct implementation of the equations of
motion. We conclude that Type IV metrics are precluded by general
covariance and this is consistent with the equations of motion.

For the remaining Type I-III configurations, in general, $g^{-1}f$
admits multiple real \textit{primary} square roots. In particular, there
is a unique \textit{principal root} which is always primary and, hence,
always consistent with general covariance. But the remaining
\textit{nonprincipal roots} can continuously degenerate into nonprimary
roots in certain regions of the configuration space where
$S^\mu{}_\nu$ will not transform as a (1,1) tensor. Hence,
requiring that the theory must always respect general covariance for 
generic field configurations, implies that  $S=\sqrt{g^{-1}f}$ must be
defined as the unique real principal square root matrix which exists
only for Type I-III metric configurations.

The above results ensure that (i) the ghost-free bimetric theory can
be defined unambiguously in terms of the principal real square root
matrix, and that (ii) the two metrics always admit compatible notions
of spacetime and compatible 3+1 decompositions, at least
locally. However, it should be emphasized that this does not yet
address the well-posedness of the initial value problem and causal
evolution, which need to be further investigated.

Finally, we comment on the issues of superluminality and causality in
bimetric theory and massive gravity. In particular, we point out that
some analyses of acausality in massive gravity that indicate the
presence of (local) closed causal curves involve configurations that
are precluded by the definition of the theory described here.

The rest of this paper is organized as follows: In Section
\ref{sec:origins} we review the origin of square root matrix in spin-2
interactions. In Section \ref{sec:theorem} we state and prove the
theorem on real square roots and interpret the resulting
classification of allowed metrics in terms of the intersections of
their null cones.  The immediate implications for compatible 3+1
decompositions are discussed in Section \ref{sec:immedimpl}.  In
Section \ref{sec:implications} we discuss the restrictions that lead
to a unique definition of the square root matrix, as the principle
root, based on general covariance. The geometric significance of the
principal square root is explained in Section \ref{sec:coord-choice}.
In Section \ref{sec:causality} we comment on causality and
superluminality issues in general and then show that massive gravity
backgrounds with local closed causal curves considered in the
literature are disallowed configurations. Finally, Section
\ref{sec:discussion} is devoted to a brief discussion of our
results. Many technical details are relegated to the appendices.

\subsection{The origin of square root matrix in spin-2 interactions}
\label{sec:origins}

On general grounds, a theory of the gravitational metric $\gmn$
interacting with another spin-2 field $\fmn$ has an action of the
form, 
\begin{align}
{\cal S}=\int\dd^4x\left[m_g^2\sqrt{-g}R_g-m^4\sqrt{-g}V(g^{-1}f) +{\cal
    L}_f(\nabla f, f)\right] \,.
\label{preBiM}
\end{align}
Here, $R_g$ is the curvature scalar of $\gmn$ and the potential $V$
can depend on the fields only through the matrix $g^{-1}f$ and its
inverse, as required by general covariance. The Lagrangian ${\cal
  L}_f$ contains the kinetic terms for $\fmn$ and may also depend on
$\gmn$. Dropping ${\cal L}_f$ and specifying $\fmn$ by hand, say,
$\fmn=\eta_{\mu\nu}$, gives massive gravity in a fixed
background. However, besides the five helicities of the massive spin-2
field $\gmn$, generally such a theory also propagates a Boulware-Deser
(BD) ghost \cite{BD}. Similar arguments also imply the existence of a
ghost in the theory with a dynamical $\fmn$. The challenge is to find
a ghost-free combination of $V$ and ${\cal L}_f$.

The historical approach to the ghost problem, briefly described below,
also sheds light on the structure of the ghost-free theory. In the
massive gravity case, the choice $\fmn=\eta_{\mu\nu}$ breaks general
covariance. The broken symmetry can be restored by the St\"{u}ckelberg
trick which amounts to replacing $\eta_{\mu\nu}$ by its coordinate
transformation,
\begin{align}
\eta_{\mu\nu}\rightarrow 
\frac{\p\phi^a}{\p x^\mu}\eta_{ab}\frac{\p\phi^b}{\p x^\nu} \,.
\label{Stuckelberg}
\end{align}
The coordinate functions $\phi^a(x)$ are the St\"{u}ckelberg fields.
On decomposing these as $\phi^a=x^a+\p^a \pi+A^a$, and taking a
certain flat space limit,
refs.\;\cite{ArkaniHamed:2002sp,Creminelli:2005qk} pointed out that a
ghost in the original massive gravity implies an unhealthy theory for
the scalar field $\pi(x)$. Namely, on setting $\gmn=\eta_{\mu\nu}$ and
$A^a(x)=0$, the replacement \eqref{Stuckelberg} amounts to,
\begin{align}
(g^{-1}\eta)^\mu_{~\nu}\to (\delta^\mu_{~\lambda} + 
\p^\mu\p_\lambda\pi)(\delta^\lambda_{~\nu}+\p^\lambda\p_\nu \pi)\,,
\label{g-eta-pi} 
\end{align}\newpage\noindent
or, in an obvious matrix notation, $g^{-1}\eta\to (\II +
\p\p\pi)^2$.  Then $V(g^{-1}\eta)$ in \eqref{preBiM} becomes a higher
derivative Lagrangian for $\pi(x)$.  If this has terms with more than
two time derivatives, it will suffer from an Ostrogradsky instability
(an unbounded nonpositive Hamiltonian) \cite{Ostrogradsky}. This
instability was correlated to the BD ghost of the original nonlinear
theory \cite{ArkaniHamed:2002sp}. To find a ghost-free theory, the
authors in \cite{Creminelli:2005qk} required that in the limit
\eqref{g-eta-pi}, $V$ should not produce terms with more than two time
derivatives of $\pi(x)$ and, hence, the derivatives must appear in the
antisymmetric combinations,
\begin{align}
V(g^{-1}\eta)\to \sum_{n=0}^4\alpha_n
\epsilon_{\mu_1\cdots\mu_n\lambda_{n+1}\cdots\lambda_4} 
\epsilon^{\nu_1\cdots\nu_n\lambda_{n+1}\cdots\lambda_4}
\p^{\mu_1}\p_{\nu_1}\pi\cdots \p^{\mu_n}\p_{\nu_n}\pi\,.
\label{Vpi}
\end{align}
This is a \textit{necessary} condition for the absence of ghost in
massive gravity. It is easy to see that this condition also uniquely
determines the nonlinear form of $V(g^{-1}\eta)$ since 
\eqref{g-eta-pi} implies the reverse replacement,
\begin{align}
\delta^\mu_\nu+\p^\mu\p_\nu\pi\to (\sqrt{g^{-1}\eta})^\mu_{~\nu} \,.
\label{reverse}
\end{align} 
Then, eliminating the $\p^\mu\p_\nu\pi$ factors through this
replacement, gives the desired potential. This was implemented in
\cite{dRGT}, based on an alternative analysis of \cite{dRG}, which
also verified the absence of the ghost at the next order. A compact
general expression, based directly on \eqref{Vpi} and valid for any
$\fmn$ is \cite{HR1103},\vspace{-1.2em}
\begin{align}
V(\sqrt{g^{-1}f})= \sum_{n=0}^4 \beta_n\, e_n(\sqrt{g^{-1}f})\,.
\label{int-pot} 
\end{align} 
The parameters $\beta_n$ are linear combinations of the $\alpha_n$ in
\eqref{Vpi}, and $e_n(S)$ are the elementary symmetric polynomials of
the eigenvalues of the matrix $S$ whose explicit forms are not needed
here. Note that the potential is uniquely determined by the necessary
condition \eqref{Vpi} of \cite{Creminelli:2005qk} for the absence of
ghost in a specific limit. Although the reasoning is valid for
$\fmn=\eta_{\mu\nu}$, we keep $\fmn$ general. An unusual feature of
$V$ is that it depends on the square root of the matrix $g^{-1}f$,
which is, in general, nonunique and not necessarily real.\footnote{The
  reasoning that leads to \eqref{reverse} assumes
  $\sqrt{(\II+\p\p\pi)^2}=\II +\p\p\pi$, analogous to $\sqrt{x^2}=x$,
  whereas, on specifying a branch, $\sqrt{x^2}=\pm|x|$. The difference
  can be ignored for small $\p\p\pi$. Early works also use a binomial
  expansion for $(\II+h)^{1/2}$, discussed in Section
  \ref{sec:discussion}.} Given the potential $V$, one can show the
absence of the BD ghost in the fully nonlinear theory
\cite{HR1106,HRS1109,HR1111,HSvS1203}.  Further work on the ghost
problem in massive gravity can be found in
\cite{Comelli:2012vz,Kluson,cedrik,Kugo:2014hja}.

Finally, the field $\fmn$ can be rendered dynamical without
reintroducing the BD ghost only through an Einstein-Hilbert action.
The ghost-free theory of two interacting spin-2 fields is then given
by \cite{HR1109},
\begin{align}
{\cal S}=\int\dd^4x\left[m_g^2\sqrt{-g}R_g-m^4\sqrt{-g}
V(\sqrt{g^{-1}f}) +m_f^2\sqrt{-f}R_f \right].
\label{BiM}
\end{align}
A Hamiltonian analysis \cite{HR1109,HR1111} shows that this theory
propagates seven modes which decompose into a massive and a massless
spin-2 fluctuation around Einstein backgrounds \cite{HSvS1515}.
However, the ghost analysis in
\cite{HR1106,HRS1109,HSvS1203,HR1109,HR1111} relies on the assumption
that the two metrics admit simultaneous 3+1 decompositions in terms of
lapse and shift variables. The complete validity of this assumption
follows from the results of the present paper. For a review of the
subsequent developments in the field, see \cite{deRrev,SvSrev}.

The $g$ and $f$ metrics can be minimally coupled (as in GR) to two
different types of matter. As long as these matter types do not
interact directly with each other, the theory remains ghost-free. If
we identify $\gmn$ as the gravitational metric, then $m_f/m_g\to
\infty$ is the massive gravity limit \cite{HR1109,Visser}, while
$m_f/m_g\to 0$ is the General Relativity limit
\cite{Akrami:2015qga}. 


\section{Theorem on real square roots and intersecting null cones}
\label{sec:theorem}

In this section we first outline two potential problems that could
render bimetric theory ill-defined, and then prove a theorem that
addresses both issues. The implications are discussed in the following 
sections.  

\subsection{Statement of the problem}
\label{sec:problems}

The previous section emphasized two main features of the ghost-free
bimetric theory \eqref{BiM}, namely, its dependence on the square root
matrix $S=\sqrt{g^{-1}f}$, and that both metrics $\gmn$ and $\fmn$
have Einstein-Hilbert terms in the action. These lead to two potential
problems:
\begin{enumerate}
\item A square root matrix function $S$ defined through $S^2=g^{-1}f$
  is not unique and not necessarily real.  It can be \textit{primary} or
  \textit{nonprimary}, furthermore, primary square roots have multiple
  branches (as many as 16 for a $4\times 4$ matrix), and the
  nonprimary ones are infinite in number (Appendix \ref{app:f-struct}
  contains an overview of matrix square roots). Thus, without a rule
  for dealing with the square root, the theory is ambiguously
  specified. To avoid this, the definition of the theory must include
  a requirement that $g^{-1}f$ has real square roots, and also a
  prescription for selecting one of the roots (similar to requiring
  invertibility of $\gmn$). Then the question is what restrictions
  this choice of $S$ imposes on $g$ and $f$.\footnote{%
    In specific cases one may be able to make sense of nonreal square 
    roots, for example allowing imaginary $S$ (corresponding to 
    $f=-c^2 g$) when the action contains only even powers of $S$
    \cite{HSvS1507}. But such procedures do not generalize to arbitrary
    $g$ and $f$  configurations that may result in nonreal $S$. Hence 
    such possibilities are not considered here.}
 
\item By definition, $\gmn$ and $\fmn$ are Lorentzian metrics of
  signature $(1,3)$. In general, two such metrics may not admit
  simultaneous proper 3+1 decompositions even locally, that is, they
  may not admit compatible notions of space and time, leading to
  potential inconsistencies outlined below.
\end{enumerate}
Recall that in General Relativity (GR) there always exist local
coordinate systems in which the metric $\gmn$ admits a \textit{proper}
3+1 decomposition, that is, with a real lapse function $N$ and a
positive definite spatial metric $g_{ij}$ (see, for example,
\cite{Gourgoulhon12}). This provides the spacelike hypersurfaces on
which initial data can be specified for local evolution.  On
restricting to globally hyperbolic spacetimes, such 3+1 decompositions
can be extended globally \cite{Choque-Bruhat, Bernal:2003jb}.  Also,
the Einstein-Hilbert action naturally splits $\gmn$ into dynamical
$g_{ij}$ and the nondynamical $N, N_i$ variables, consistent with the
3+1 decomposition. The existence of a global time allows for a
Hamiltonian formulation of the theory \cite{Dirac,ADM60}.

The two Einstein-Hilbert terms in the bimetric action \eqref{BiM}
indicate that, for each metric, a 3+1 decomposition is still relevant
for isolating its dynamical content. The absence of compatible 3+1
splits for the two metrics then leads to inconsistent evolution
equations, even locally. For example, when such metrics couple to
their respective matter sectors, there may not exist common
hypersurfaces for specifying initial data for both sectors. Hence,
consistent time evolution requires that the metrics admit compatible
3+1 decompositions (although this alone is not enough for the Cauchy
problem to be well-posed). Note that in massive gravity, where the $f$
metric is nondynamical, the existence of compatible 3+1 decompositions
is not a requirement, unless the theory is regarded as a bimetric
limit. 

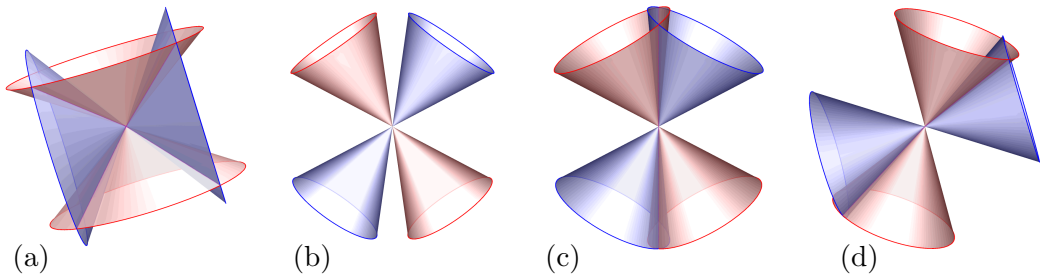
\begin{figure}[h]
\centering{}\hspace{2mm}\begin{tikzpicture}[x=1mm,y=1mm]
   \node[anchor=south west, inner sep=0] at (0,0) {
      \includegraphics[width=140mm]{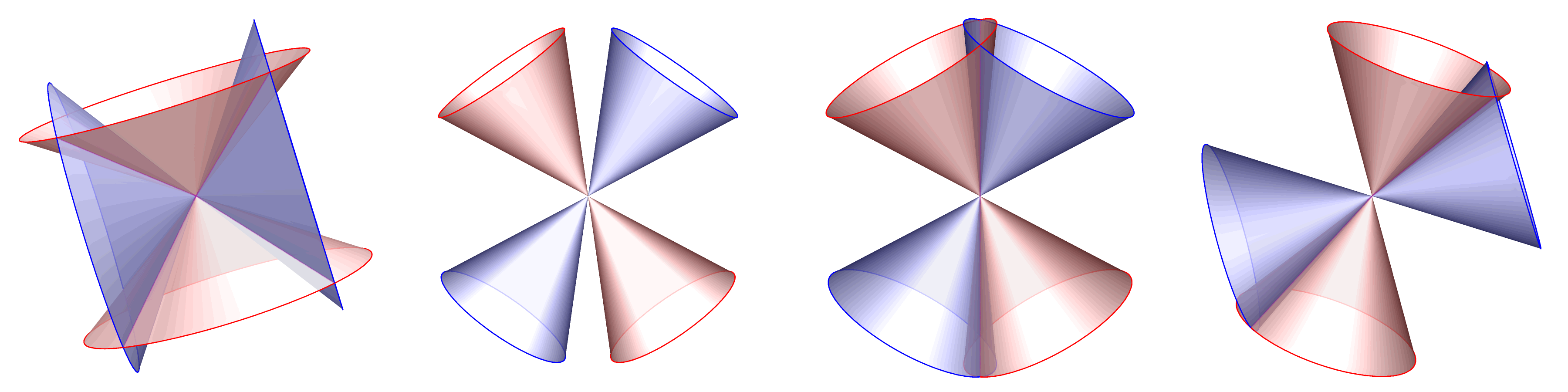}
      };
   \node[] at (0*35+5,0) {(a)};
   \node[] at (1*35+7,0) {(b)};
   \node[] at (2*35+5,0) {(c)};
   \node[] at (3*35+9,0) {(d)};
\end{tikzpicture}\vspace{-4mm}\protect\caption{\label{fig:bad-adm}
Some possible intersections of two null cones}
\end{figure}

The situation is easily illustrated in terms of the null cones of the
two metrics, with a few examples shown in Figure \ref{fig:bad-adm} (a
longer list is given later).\footnote{We use the term of null cone
  instead of light cone. To be specific, at a point $p$ on a
  spacetime manifold $M$, a null cone is a subset of the tangent space
  at $p\in M$.  Relative to some metric, it is a set of all null
  vectors at $p$, excluding the zero vector.  The \textit{interior} of
  a null cone is a set of all timelike vectors while its
  \textit{exterior} is a set of all spacelike vectors at $p$.  A
  \textit{surface element} through $p$ is called \textit{spacelike} if
  and only if its normal is timelike, and \textit{null} if and only if
  its normal is null. A \textit{spacelike surface} is one which is
  spacelike everywhere. A spacelike surface element through $p$ does
  not intersect the null cone, i.e., it contains only the zero vector
  and separates the two parts of the null cone into past and future.}

The null cones in case (a) have common timelike directions within
their intersection, but no common spacelike hypersurfaces, hence they
do not admit simultaneous proper 3+1 decompositions.  In cases (b) and
(c), the two metrics admit compatible 3+1 decompositions with real
lapses and common spatial hypersurfaces. However, in case (b) the choice of
time and space directions is not unique as the time axis can be chosen
either vertically or horizontally. As a result, the notion of time
orientation is not unique (an upward time axis aligns with the future
of $g$ and $f$ cones, whereas a horizontal time axis could align with
the future of $g$ and past of $f$). Only in case (c) the null cones
are interlocked along common timelike directions within their
intersection and have the same relative time orientation. Case (d) is
a coordinate transformation of (c) and admits proper 3+1 splits for
correctly chosen spatial hypersurfaces. In the following it is enough
to consider metrics up to diffeomorphisms, so cases (c) and (d) are
not distinguished.\footnote{Besides the tangent space null cones, given
  by $\gmn v^\mu v^\nu=0$, one may also consider cotangent, or
  momentum space null cones, $g^{\mu\nu} k_\mu k_\nu=0$. The momentum
  space null cones corresponding to Figure \ref{fig:bad-adm}(a) have
  the topology of case (b) and vice versa. Only in the intersecting
  case (c), the tangent and the cotangent space null cones have
  similar topology.\label{fn:cotangent}}


\subsection{Definition of causally coupled and null coupled metrics}
\label{def:coupled}

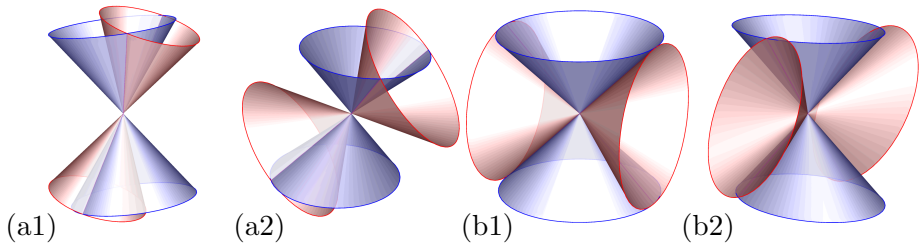
\begin{figure}[h]
\centering{}\hspace{2mm}\begin{tikzpicture}[x=1mm,y=1mm]
    \node[anchor=south west, inner sep=0] at (0,0) {
      \includegraphics[width=120mm]{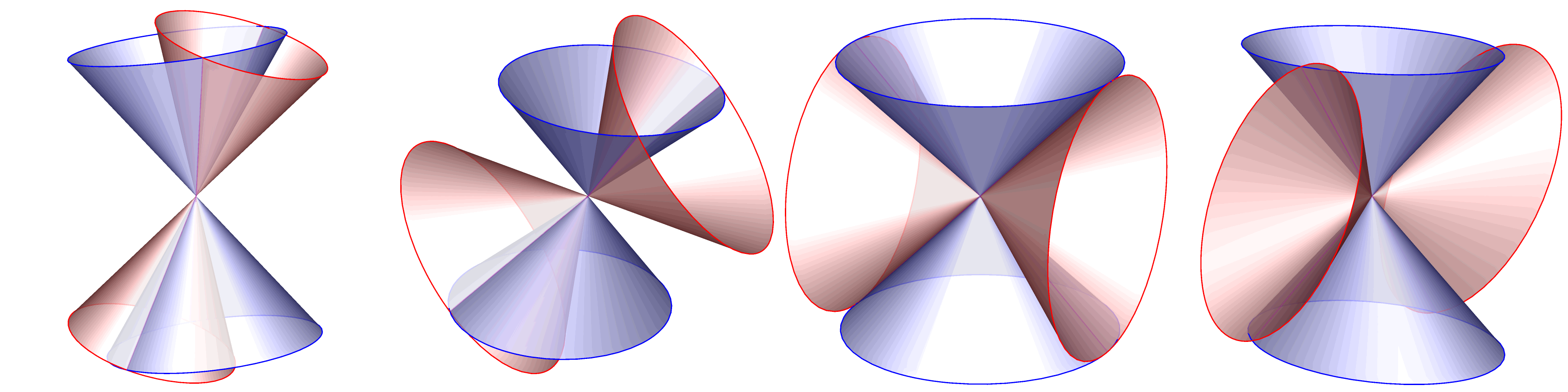}
      };
   \node[] at (0*30+3,0) {(a1)};
   \node[] at (1*30+3,0) {(a2)};
   \node[] at (2*30+3,0) {(b1)};
   \node[] at (3*30+2,0) {(b2)};
\end{tikzpicture}\vspace{-4mm}\protect\caption{\label{fig:coupled}(a)
  Causally coupled, and (b) Null coupled cones}     
\end{figure}

We introduce some terminology to distinguish two types of null cone
intersections. Let $M$ be a manifold equipped with two arbitrary
Lorentzian metric tensors $f$ and $g$, each of signature
$(1,n)$. Consider their null cones at some point $p\in M$. We say that
$f$ and $g$ (or their null cones) are \textit{causally coupled at} $p$
if and only if there exists, through $p$, a common timelike vector and
a common spacelike surface element relative to both $f$ and
$g$. Examples are shown in Figure \ref{fig:coupled}(a). We say that
$f$ and $g$ (or their null cones) are \textit{null coupled at} $p$ if
and only if there neither exists a common timelike vector nor a common
spacelike surface element relative to both $f$ and $g$ at $p$,
i.e., when the null cones touch along two null directions.
Examples are shown in Figure \ref{fig:coupled}(b).

\begin{figure}[h]
\centering{}\hspace{2mm}\begin{tikzpicture}[x=1mm,y=1mm]
    \node[anchor=south west, inner sep=0] at (0,0) {
      \includegraphics[width=120mm]{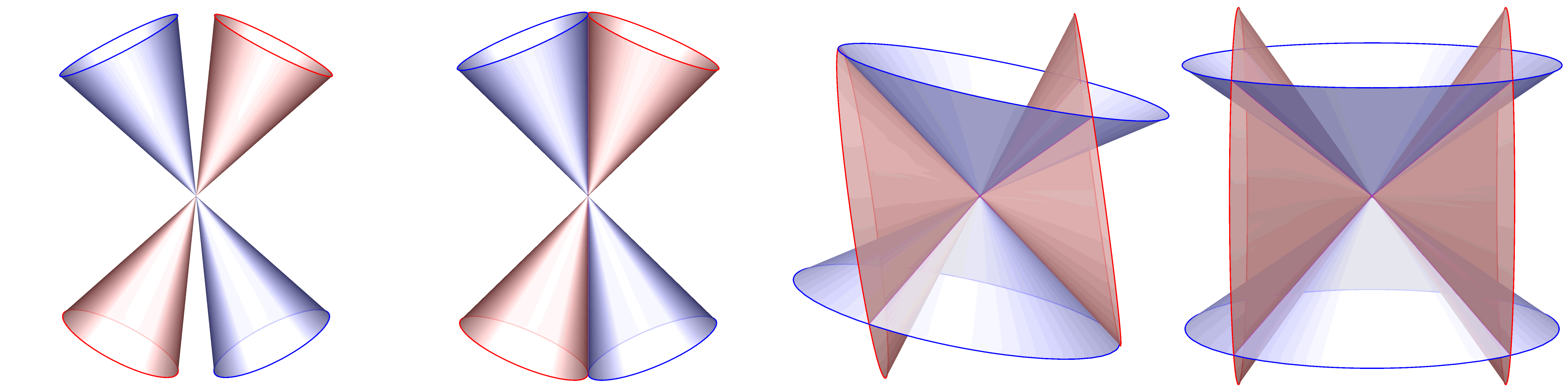}
      };
   \node[] at (0*30+4,0) {(a)};
   \node[] at (1*30+3,0) {(b)};
   \node[] at (2*30+4,0) {(c)};
   \node[] at (3*30+0,0) {(d)};
\end{tikzpicture}\vspace{-4mm}\protect\caption{\label{fig:decoupled}
Null cones that are neither causally nor null coupled} 
\end{figure}
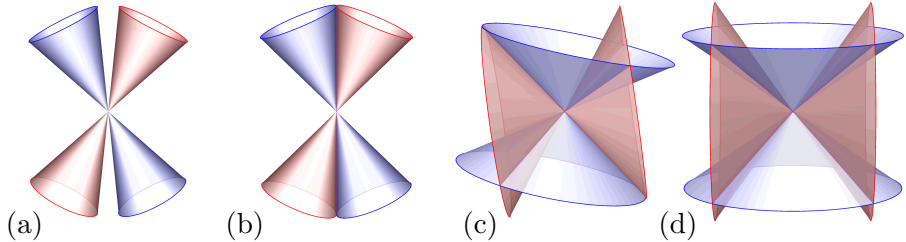

For comparison, Figure \ref{fig:decoupled} shows typical cases when
$f$ and $g$ are neither causally coupled nor null coupled.
The above definitions are local statements about the topology of null
cones, independent of the coordinate system. Recall that the interior
(and the exterior) of a null cone is an open set, and that the
intersection of two open sets is an open set. When the $f$ and $g$
null cones are causally coupled, their interiors and exteriors
intersect so that one can always find an inner cone inside the
intersection of their interiors. Similarly one can always find an
outer cone the interior of which contains the null cones of both $f$
and $g$. A special case is when one null cone is contained within the
other.

\subsection{Statement and proof of the theorem}
\label{the:interlock}

We now state the main result of this paper as a theorem and provide a
proof.  
\begin{thm*}
Let $f$ and $g$ be two regular Lorentzian metric tensors. A real
square root of $A\equiv g^{-1}f$ exists if and only if $f$ and $g$ are
either causally coupled or null coupled.
\end{thm*}
As described in Appendix \ref{app:f-struct}, a matrix can have several
real primary square roots and an infinite number of nonprimary
roots. The theorem applies regardless of which one of these is chosen,
but it also provides a classification of the real roots. Then, as
explained in the next section, the requirement of general covariance,
leads to a unique choice for the square root.

Here we prove that the reality of $\sqrt{g^{-1}f} $ is sufficient for
$f$ and $g$ to be causally coupled or null coupled (this is the
``\textit{only if}'' part of the theorem). The proof of the necessary
condition is less instructive and is relegated to Appendix
\ref{app:converse}. The strategy of the sufficiency proof is first to
state the conditions under which $A$ admits real square roots and
then, for $A=g^{-1}f$, to find the corresponding restrictions on $g$
and $f$ using a theorem on the \textit{canonical form of a pair of
  matrices}.  We work \textit{locally} at some point $p$ of a manifold
of finite dimension $d=1+n$, with 1 time and $n\ge1$ space dimensions.
$\fmn$ and $\gmn$ are nonsingular real symmetric tensors.  Locally,
congruences $Z^{\tr}gZ$ and $Z^{\tr}fZ$, with real $Z$, are equivalent
to coordinate transformations at $p$.

\subsubsection{Reality of the square root matrix}\label{sec:realSQR}

All eigenvalues of $g$ and $f$ are real, but the matrix $A=g^{-1}f$ is
not symmetric and can have complex eigenvalues. Let us denote its real
eigenvalues by $\lambda_k$, $1\le k\le q$, and the complex ones, which
come in conjugate pairs, by $a_k\pm \ii b_k$, $k\ge q+1$. $A$ may not
be diagonalizable, but it always has a Jordan normal form (see section
\ref{sec:pair}). A real square root $\sqrt{A}$ exists only in the
following two cases (see theorems~1.23, 1.26 and 1.29 in 
\cite{Higham08}, see also \cite{Higham87,Horn94}):
\begin{enumerate}
\item No eigenvalue of $A$ lies on the negative real axis
  $\mathbb{R}^{-}$. This restricts real eigenvalues to
  $\lambda_{k}>0$, $1\le k\le q$, but there is no restriction on
  complex eigenvalues.
\item The matrix $A$ has real negative eigenvalues $\lambda_{k}<0$,
  but each such eigenvalue appears in an even number of Jordan blocks
  of the same size.
\end{enumerate}
$A$ can be put in Jordan normal form, $A=ZA_\eJ\,Z^{-1}$, by some
appropriate transformation $Z$. Then, by definition,
$\sqrt{A}=Z\sqrt{A_\eJ}\,Z^{-1}$, where the square roots $\sqrt{A_\eJ}$
can be explicitly computed by standard methods (see Appendix
\ref{app:f-struct}). $\sqrt{A}$ is a real matrix only in the above two
cases even when it has complex eigenvalues, hence the action and the
equations of motion will be real. Below we convert the above
conditions on $A$ into restrictions on $g$ and $f$ and express them in
terms of null cone intersections.

\subsubsection{Canonical form for a pair of symmetric matrices}
\label{sec:pair}

The transformation $Z$ that converts $A$ to its Jordan normal form
$A=ZA_\eJ\,Z^{-1}$ is not unique. For $A=g^{-1}f$, the arbitrariness
in $Z$ can be used to put $g$ and $f$ in a specific form, $Z^{\tr}gZ$,
$Z^{\tr}fZ$, determined by the structure of $A_\eJ$. This is
accomplished by a theorem on \textit{canonical pair form}, a proof of
which can be found in \cite{Uhlig73}.\footnote{Many authors, starting
  with Weierstrass and Kronecker, have worked on the question of
  canonical forms for a pair of real symmetric matrices. For a
  historical and mathematical survey see \cite{Uhlig76,Uhlig79}, also
  \cite{Gant59v2,Hong86,Horn90}.} Then, conditions on the Jordan
normal form of $g^{-1}f$ can be converted into restrictions on $g$ and
$f$.

The theorem on \textit{canonical pair form} states that, given
nonsingular real symmetric matrices $g$ and $f$, there always exists a
real, nonsingular transformation $Z$ that converts $g^{-1}f$ to its
Jordan normal form and, at the same time, puts $g$ and $f$ in specific
forms, 
\begin{align}
Z^{-1}g^{-1}fZ & =\diag(J_{1},\dots,J_{q},C_{q+1},\dots,C_{p}),
\label{eq:cpt-3}\\
Z^{\tr}gZ & =\diag(\epsilon_{1}E_{1},\dots,\epsilon_{q}E_{q},\,
E_{q+1},\dots,E_{p}),\label{eq:cpt-1}\\ 
Z^{\tr}fZ & =\diag(\epsilon_{1}E_{1}J_{1},\dots,\epsilon_{q}
E_{q}J_{q},\, E_{q+1}C_{q+1},\dots,E_{p}C_{p}).\label{eq:cpt-2}  
\end{align}
The first line is simply the Jordan normal form of $g^{-1}f$ where  
$J_{k}$ are the $n_k\times n_k$ Jordan blocks (for $1\le k\le q$)
corresponding to the real eigenvalues $\lambda_{k}$ of $A$,
while $C_{k}$ are the $n_k\times n_k$ real Jordan blocks (for $k\ge
q+1$ and even $n_{k}$) corresponding to the pairs of complex conjugate
eigenvalues $a_{k}\pm\ii b_{k}$ of $A$, 
\begin{equation}
J_{k}\equiv
\begin{pmatrix}
\lambda_{k} & 1\\[-0.5em]
 & \lambda_{k} & \ddots\\[-0.4em]
 &  & \ddots & 1\\[-0.1em]
 &  &  & \lambda_{k}
\end{pmatrix}\!,\quad 
C_{k}\equiv
\begin{pmatrix}\Lambda_{k} & I_{2}\\[-0.5em]
 & \Lambda_{k} & \ddots\\[-0.4em]
 &  & \ddots & I_{2}\\[-0.1em]
 &  &  & \Lambda_{k}
\end{pmatrix}\!,\ \Lambda_{k}\equiv
\begin{pmatrix}a_{k} & -b_{k}\\
b_{k} & a_{k}
\end{pmatrix}\!,\ I_{2}\equiv
\begin{pmatrix}1 & 0\\
0 & 1
\end{pmatrix}\!.\label{eq:rjb1}
\end{equation}
The transformation $Z$ in \eqref{eq:cpt-3} is not uniquely fixed since
there are many matrices that commute with Jordan form on the right
hand side.  The freedom in choosing $Z$ is just enough to transform
$g$ as in equation \eqref{eq:cpt-1}, where $\epsilon_{k}\in\{\pm1\}$
are possible signs related to the \textit{signature} of $g$, and
$E_{k}=E_{k}^{-1}$ is the $n_{k}\times n_{k}$ exchange (or reversal)
matrix,
\begin{equation}
E_{k}\equiv\begin{pmatrix} &  & 1\\[-0.5em]
 & \iddots\\[-0.5em] 1 \end{pmatrix}\!.
\label{eq:Ek}
\end{equation}
Here, $\dim E_{k}=\dim J_{k}$ for $k\le q$ and $\dim E_{k}=\dim C_{k}$
for $k>q$. Finally, given $g^{-1}f$ and $g$, $f$ can be computed to be 
\eqref{eq:cpt-2}. While all eigenvalues of $f$ and $g$ are necessarily
real, this may not hold for $g^{-1}f$. Nonetheless, all matrices in
the theorem are real, including $Z$.

We again emphasize that, at any point $p$, the congruence or similarity
transformation $Z$ can be interpreted as a local coordinate
transformation. In fact, since $Z^{\tr}gZ$ in \eqref{eq:cpt-1} is the
Lorentz frame metric $\eta$, at most in a nondiagonal form, $Z$ coincides
with the vielbein of the metric $g$ in such a Lorentz basis.

\subsubsection{Proof by enumeration}
\label{sec:proofenum}

Using the above decomposition, we now enumerate all possible types of
matrices $g$ and $f$ subject to the conditions that the real
eigenvalues $\lambda_k$ of $g^{-1}f$ are restricted by the reality of
$\sqrt{g^{-1}f}$, as specified in subsection \ref{sec:realSQR}, while 
the complex eigenvalues are unrestricted. Also the following obvious 
restrictions apply: 
\begin{enumerate}
\item Let $\rho^{-}$ ($\rho^{+}$) denote the number of negative
  (positive) eigenvalues of a matrix. Then,
\begin{equation}
  \rho^{-}(g)=\rho^{-}(f)=
    1,\qquad\rho^{+}(f)=\rho^{+}(g)=n,\qquad d=n+1\ge2\;.
\end{equation}
\item The number and type of Jordan blocks may vary, yet they always
  satisfy $\dim E_{k}=\dim J_{k}$ for $k\le q$, corresponding to real
  eigenvalues $\lambda_k$, and $\dim E_{k}=\dim C_{k}$ for $k>q$,
  corresponding to the complex eigenvalues $a_k\pm \ii b_k$.
\end{enumerate}
The signature of $g$, decomposed as in \eqref{eq:cpt-1}, imposes 
the following condition on the $E_k$'s, 
\begin{align}
1 & =\rho^{-}(g)=\sum_{1\le k\le q}\rho^{-}(\epsilon_{k}E_{k})+
\sum_{q<k\le p}\rho^{-}(E_{k}),
\end{align}
where, for the matrices $E_k$ given by \eqref{eq:Ek} one has, 
\begin{equation}
\rho^{-}(E_{k})=\left\lfloor{\tfrac{1}{2}}n_{k}\right\rfloor, 
\quad\rho^{+}(E_{k})=\left\lfloor {\tfrac{1}{2}}
\left(n_{k}+1\right)\right\rfloor,\quad n_{k}=\dim E_{k}.
\end{equation}
Thus, up to permutations of the blocks, $Z^{\tr}gZ$ can only have
a finite number of possible forms which restrict $Z^{\tr}fZ$ accordingly,
given the Jordan form of $g^{-1}f$. Denoting $Z^{\tr}gZ$ and $Z^{\tr}fZ$
as $g$ and $f$ for brevity, all the possibilities are enumerated below. We
focus on $d=4$ for definiteness, but the results hold for any dimension.
\vspace{0.5em}

\underline{\textit{No $E_k$ with $n_k>1$}:} Here,
$g=\diag(\epsilon_1,\epsilon_2,\epsilon_3,\epsilon_4)$ has only
one negative $\epsilon_k$, which we can assume to be
$\epsilon_1=-1$.  Then, $\dim J_k =1$ implies
$g^{-1}f=\diag(\lambda_1,\lambda_2,\lambda_3,\lambda_4)$ and
$f=\diag(-\lambda_1,\lambda_2,\lambda_3,\lambda_4)$, with real
$\lambda_k$.  The reality of the square root requires the following:
(i) Either all $\lambda_k>0$, or, (ii) there exists one pair of
negative eigenvalues $\lambda_i=\lambda_j<0$ with other
$\lambda_k>0$, or (iii) more than one pair of type
$\lambda_i=\lambda_j<0$, and other $\lambda_k>0$. Finally,
$\rho^-(f)=1$ allows at most one negative pair ruling out (iii) and
also requires the negative pair to include $\lambda_1$. Thus only two
types of configurations are allowed (with $\lambda,\lambda_k>0$), 
\begin{alignat}{3}
 & \textbf{Type I}:
 & \qquad & g = \diag(-1,1,1,1), 
 & \qquad & f = \diag(-\lambda_1,\lambda_2,\lambda_3,\lambda_4),
\label{type1} \\
 & \textbf{Type IV}: 
 & \qquad & g = \diag(-1,1,1,1), 
 & \qquad & f = \diag(\lambda,-\lambda,\lambda_3,\lambda_4).
\label{typeIV}
\end{alignat}

\underline{\textit{One $E_k$ with $n_k=2$}:} In this case,
$g=\diag(\epsilon_1 E_1, 1,1)$ with $\dim E_1=2$. When
$\epsilon_1=+1$, $g^{-1}f$ contains either a 2$\times$2 $J_1$ block
or a 2$\times$2 $C_1$ block, with the remaining $J_k$ being one
dimensional. Then the reality of the square root and $\rho^-(f)=1$
leads to ($\lambda,\lambda_k>0, b\neq 0$),
\begin{alignat}{3}
 & \textbf{Type IIa}: 
 & \qquad & g = 
   \diag(\begin{pmatrixc}
     0&1\\[-0.2em]1&0\end{pmatrixc}\!,1,1 ), 
 & \qquad & f = 
   \diag(\begin{pmatrixc} 0&\lambda\\[-0.2em]\lambda
   &1 \end{pmatrixc}\!,\lambda_2,\lambda_3),  
\label{type2a}  \\ 
 & \textbf{Type IIb}: 
 & \qquad 
 & g = \diag(\begin{pmatrixc} 0 & 1 \\[-0.2em]1 & 0 \end{pmatrixc}\!,1,1), 
 & \qquad 
 & f = \diag(\begin{pmatrixr}b & a \\[-0.2em] a &
     -b \end{pmatrixr}\!,\lambda_2,\lambda_3).
\label{type2b}
\end{alignat}
The case with $\epsilon_1=-1$ corresponds to Type IIa with the sign of
the 2$\times$2 block reversed. When $\epsilon_1=1$ the $f$
null cone is inside the $g$ null cone and \textit{vice versa} for 
$\epsilon_1=-1$. But except for the interchange of the metrics, 
the topology of the intersection remains unchanged.

\underline{\textit{One $E_k$ with $n_k=3$}:} Here, 
the only allowed configuration is with 3$\times$3 blocks $E_1$, $J_1$, 
\begin{alignat}{3}
 & \textbf{Type III}: 
 & \qquad & g = \diag(E_1,1),
 & \qquad & f = \diag(E_1 J_1,\lambda_2).
\label{type3}
\end{alignat}
\vspace{0.2em}

In all the above cases $g^{-1}$ has the same matrix as $g$;
hence, reintroducing the factors of $Z$, the matrix of $Z^{-1}g^{-1}fZ$
can be easily computed. The enumeration \eqref{type1}-\eqref{type2b}
is valid for all spatial dimensions $n\ge1$, while the presence
of Type III is restricted to $n\ge2$.\footnote{In dimensions higher
  than $d=2$, only a common positive definite part of the two metrics 
  is affected by the additional positive eigenvalues.}
Finally, the explicit construction of a common timelike vector and
a common spacelike surface element for each Type is given in Appendix
\ref{app:proof}, completing the sufficiency proof.
A comprehensive overview of the allowed configurations 
is given in Table \ref{tab:segre} with the Segre characteristics of 
$g^{-1}f$ shown in the second column.\footnote{%
  The Segre characteristic is a descending list of integers 
  that correspond to the sizes of the blocks in a Jordan normal form
  where complex blocks are designated by $z\bar{z}$ instead.
  The integers representing submatrices with the same eigenvalue 
  are listed together in parentheses;
  for example, $[(21)1]$ denotes a derogatory Type IIa, $[211]$, where
  $\lambda=\lambda_2\ne\lambda_3$.
}\vspace{-0.5em}

\begin{table}[h]
  \caption{Overview of all possible metric configurations  
  for $d=4$. \label{tab:segre}}\vspace{-0.3em}
  \begin{center}
    \bgroup\renewcommand{\arraystretch}{1.5}
    \begin{tabular}{ccccc} \hline\hline
      Type & Segre char. & $\diag(g)$ & $\diag(f)$ & $\diag(g^{-1}f)$ 
      \\ \hline
      \textbf{I} & $[1111]$ 
      & $(-1,1,1,1)$ 
      & $(-\lambda_1,\lambda_2,\lambda_3,\lambda_4)$
      & $(\lambda_1,\lambda_2,\lambda_3,\lambda_4)$
      \\
      \textbf{IIa} & $[211]$ 
      & $(\pm\!
        \begin{pmatrixc} 0&1\\[-0.2em]1&0\end{pmatrixc}\!,1,1 )$
      & $(\pm\!
        \begin{pmatrixc} 0&\lambda\\[-0.2em]\lambda & 1 \end{pmatrixc}\!,
        \lambda_2,\lambda_3)$
      & $(\begin{pmatrixc} \lambda&1\\[-0.2em]0 & \lambda \end{pmatrixc}\!,
      \lambda_2,\lambda_3)$
      \\[0.8em]
      \textbf{IIb} & $[z\bar{z}11]$ 
      & $(\pm\!
        \begin{pmatrixc} 0 & 1 \\[-0.2em]1 & 0 \end{pmatrixc}\!,1,1)$ 
      & $(\pm\!
        \begin{pmatrixr}b & a \\[-0.2em] a & -b \end{pmatrixr}\!,
        \lambda_2,\lambda_3)$ 
      & $(
        \begin{pmatrixr}a & -b \\[-0.2em] b & a \end{pmatrixr}\!,
        \lambda_2,\lambda_3)$ 
      \\[0.8em]
      \textbf{III} & $[31]$ 
      & $(\begin{pmatrixc} 0&0&1\\[-0.2em]
      0 & 1 & 0\\[-0.2em]
      1 & 0 & 0 \end{pmatrixc}\!,1)$
      & $(\begin{pmatrixc} 0&0&\lambda\\[-0.2em]
      0 & \lambda &1\\[-0.2em]
      \lambda & 1 & 0 \end{pmatrixc}\!,\lambda_2)$
      & $(\begin{pmatrixc} \lambda&1&0\\[-0.2em]
      0 & \lambda &1\\[-0.2em]
      0 & 0 & \lambda \end{pmatrixc}\!,\lambda_2)$
      \\
      \textbf{IV} & $[(11)11]$ 
      & $(-1,1,1,1)$ 
      & $(\lambda,-\lambda,\lambda_2,\lambda_3)$
      & $(-\lambda,-\lambda,\lambda_2,\lambda_3)$
      \\ \hline\hline
    \end{tabular}
    \egroup
  \end{center}
\end{table}\vspace{-1em}

\begin{figure}[h]
\centering{}\hspace{2mm}\begin{tikzpicture}[x=1mm,y=1mm]
    \node[anchor=south west, inner sep=0] at (0,4) { 
      \includegraphics[width=150mm]{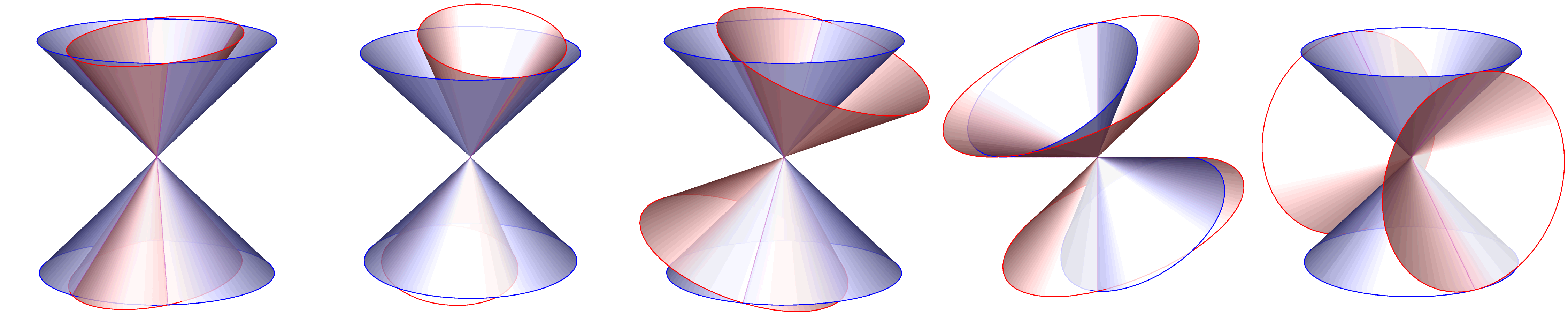}
      };
   \node[] at (0*30+15,0) {Type I};
   \node[] at (1*30+15,0) {Type IIa};
   \node[] at (2*30+15,0) {Type IIb};
   \node[] at (3*30+15,0) {Type III};
   \node[] at (4*30+15,0) {Type IV};
\end{tikzpicture}\vspace{-3mm}\protect\caption{\label{fig:jbs}Null cones
  corresponding to different allowed $g$ and $f$ metrics.}
\end{figure}
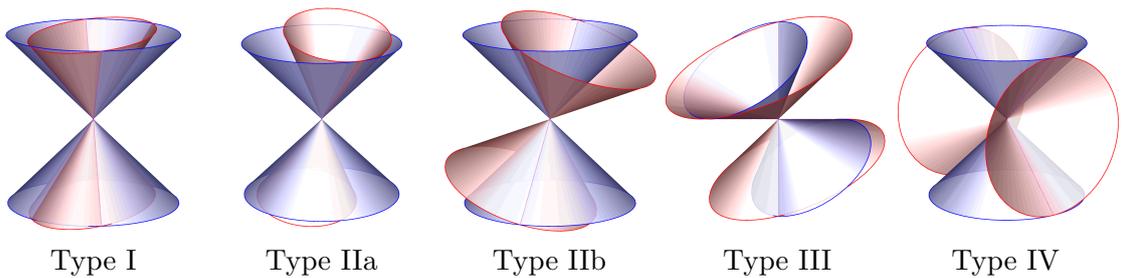

The possible configurations of the metrics are easily visualized in terms
of the intersections of their null cones. The $Z$ transformation deforms
the cones, but does not change the nature of their intersections. For each
Type, the null cones are depicted, by example, in Figure \ref{fig:jbs} in
a convenient coordinate system.

From Figure \ref{fig:jbs} it is obvious that for the configuration
Types I-III, the null cones are \textit{causally coupled}, while in Type
IV, they are \textit{null coupled} (using the terminology of section
\ref{def:coupled}). The permitted configuration types I-IV correspond
to the null cone intersections depicted in Figure \ref{fig:jbs} for
all allowed values of the parameters $\lambda,\lambda_k,a,b$. This is
easy to verify analytically using the equations in Appendix
\ref{app:proof}, or simply by plotting the null cones for different
parameter values.

Examples of the above results are the two metric ans\"atze considered
in \cite{Baccetti:2012ge}, for which the square root is real and the
null cones belong to subsets of Type IIa and Type I.  Note that, of
the allowed configurations in 4 dimensions, only Type I and Type IIb
can involve the full set of 20 components in $f$ and $g$, with 16
components in $Z$ and 4 components in $Z^\tr f Z$. The remaining
configurations cannot realize the full number of field components.


\subsection{Implications of the theorem}
\label{sec:immedimpl}

The immediate implication of the above theorem is that the reality of
bimetric equations of motion restricts the metrics to be either
causally coupled (Type I-III) or null coupled (Type IV), as depicted
in Figure \ref{fig:jbs}. Causally coupled metrics admit simultaneous
\textit{proper} 3+1 decompositions, hence, the equations of motion can
be recast as time evolution equations, as a first step towards
formulating the Cauchy problem. The interlocking of the null cones
ensures that if the spacetime manifold is time-orientable with respect
to the metric $g$, then it is also time-orientable with respect to $f$
and \textit{vice versa}.

Null coupled (Type IV) metrics do not admit simultaneous proper 3+1
decompositions and would be problematic if allowed by the theory.
Some of these issues will be discussed later. The simplest realization
of such configurations is $g=\diag(-1,1,1,1)$,
$f=\diag(1,-1,1,1)$. In general relativity, the metric also admits
a d'Inverno-Stachel-Smallwood (DSS) 2+2 decomposition
\cite{DSS1,DSS2}, and initial data can be specified on a pair of null
hypersurfaces. Since null coupled metrics share \textit{two} null
hyperplanes, they admit simultaneous 2+2 decompositions but still
remain problematic. The next section shows that Type IV configurations
can and should be consistently discarded.

Metrics that are neither causally coupled nor null coupled, as in
Figure \ref{fig:decoupled}, result in complex equations of motion and
are disallowed. If such configurations occurred, time-orientability
with respect to one metric would not ensure the same for the other
metric. Note that, although the configuration in Figure
\ref{fig:decoupled}(b) seems to be a limit of the causally coupled
Type IIb cones in Figure \ref{fig:jbs}, with the intersection shrunk
to zero, in reality, it is not obtainable from any configuration with
real $\sqrt{g^{-1}f}$. Instead, it arises from metrics parameterized
as Type IIa in \eqref{type2a}, but with $\lambda<0$, corresponding to
a nonreal $\sqrt{g^{-1}f}$.

\section{Unique choice of the square root matrix}
\label{sec:implications}
The reality condition alone does not determine square root matrix
$S=\sqrt{g^{-1}f}$ uniquely. In this section we show that general
covariance uniquely restrict $S$ to the \textit{primary square root on
  the principal branch}, ruling out, in particular, the null coupled
(type IV) configurations. This is shown to be consistent with the
equations of motion, insuring that the two metrics admit compatible
notions of space and time.

\subsection{Choice of square root matrix from general covariance}
\label{sec:h} 
The bimetric equations of motion have the form,\footnote{%
  For an explicit derivation of the equations, see, for example, 
  the appendix in \cite{HSvS1212}. }
\begin{align}
(G^g)^\mu_{~\nu}+\sum_{n=0}^3 c^g_n\,(S^n)^\mu_{~\nu}=0\,,\qquad
(G^f)^\mu_{~\nu}+\sum_{n=0}^3 c^f_n\, (S^n)^\mu_{~\nu}=0
\end{align}
where $G^{g}$ and $G^{f}$ denote the Einstein tensors for $g$ and $f$
metrics, respectively, and the coefficients $c_n^{g}$ and $c_n^{f}$
are functions of $\Tr(S^m)$. Under general coordinate transformations
$x^\mu\rightarrow \tilde x^\mu$, the matrix
$A^\mu_{~\nu}=g^{\mu\lambda}f_{\lambda\nu}$ transforms as a (1,1)
tensor to $\tilde A(\tilde x)=Q^{-1}A(x)Q$, where $Q^\mu_{~\nu}=\p
x^\mu/\p \tilde x^\nu$. The bimetric equations will transform
covariantly only if $S=\sqrt{A}$ also transforms as a (1,1) tensor,
which requires that,
\begin{align}
\sqrt{Q^{-1} A Q}= Q^{-1}\sqrt{A} Q \,.
\end{align}
Nonprimary square roots do not satisfy this property and hence can be
discarded by the requirement of general covariance. We now show that
this leads to a unique choice for the square root matrix.

The nature of the square root of $A$ is determined by its Jordan
normal form $A_\eJ$ (see Appendix \ref{app:f-struct}). First, note that
for Type IV configurations,
$A_\eJ=\diag(-\lambda,-\lambda,\lambda_3,\lambda_4)$ and its primary
square roots contain the 2$\times$2 blocks
$\pm\diag(i\lambda^{1/2}, i\lambda^{1/2})$. These are nonreal in a 
real basis and are ruled out by the reality condition. On the other
hand, square roots containing the 2$\times$2 blocks
$\pm\diag(i\lambda^{1/2},-i\lambda^{1/2})$ have a real form in an
appropriate basis which is the reason why Type IV metrics are not
ruled out by the reality condition. However, the latter roots are
nonprimary and violate general covariance. Thus, Type IV
configurations do not admit real square roots that preserve general
covariance.

In the remaining Type I-III configurations, $A$ has no negative real
eigenvalue, hence it admits multiple real primary square roots. Let us
denote its distinct eigenvalues by $\lambda_i$ and the corresponding
Jordan blocks by $J_i$ (there are 2 to 4 such blocks depending on the
configuration). Then, $A_\eJ=\diag(J_1,\cdots,J_s)$.  Assuming that
the eigenvalues in different Jordan blocks are distinct, the different
primary square roots are,
\begin{align}
\sqrt{A_\eJ}=\diag(\pm J_1^{1/2},\cdots,\pm J_s^{1/2})\,.
\label{eq:AJ}
\end{align}
Here, $J_i^{1/2}$ are given by eq.\,\eqref{eq:SqrtJ} and the $\pm$
signs are chosen independently. Selecting the $+$ sign for all blocks
gives the principal square root. (For the purpose of this argument we
do not distinguish between the principal roots $\sqrt{A}$ and
$-\sqrt{A}$ as the extra overall sign can be absorbed in the $\beta_n$
parameters of the theory.) Other combinations of the signs produce the
remaining primary square roots as long as the $\lambda_i$ remain
distinct. However, it is possible that due to symmetries, or during
the evolution, two eigenvalues collapse to the same value, say,
$\lambda_k=\lambda_l$, without producing a discontinuity in the square
root. Then, if the corresponding $J_k^{1/2}$ and $J_l^{1/2}$ in
eq.\,\eqref{eq:AJ} have opposite signs, the square root becomes
nonprimary. Thus, nonprincipal square roots become nonprimary whenever
eigenvalues in blocks with different signs in \eqref{eq:AJ} become
equal. Only the principal square root avoids becoming nonprimary for
the allowed range of $\lambda_i$, since all blocks $J_i^{1/2}$ are
guaranteed to have the same sign.

To summarize, requiring reality and general covariance of the
equations restricts $\sqrt{g^{-1}f}$ uniquely to be the principal real
square root which is always a primary matrix function. In particular,
Type IV configurations are disallowed since their only real square
roots are nonprimary. The allowed configurations are the causally
coupled metrics (Type I-III) which always admit compatible 3+1
decompositions. In subsection \ref{sec:noIV} we will show that
disallowing Type IV metrics is consistent with the equations of motion
and is, in fact, also required by the variational principle.

\subsection{Geometric significance of the principal root} 
\label{sec:coord-choice}

\begin{figure}[h]
\centering{}\hspace{2mm}\begin{tikzpicture}[x=1mm,y=1mm]
    \node[anchor=south west, inner sep=0] at (0,4) {
      \includegraphics[width=150mm]{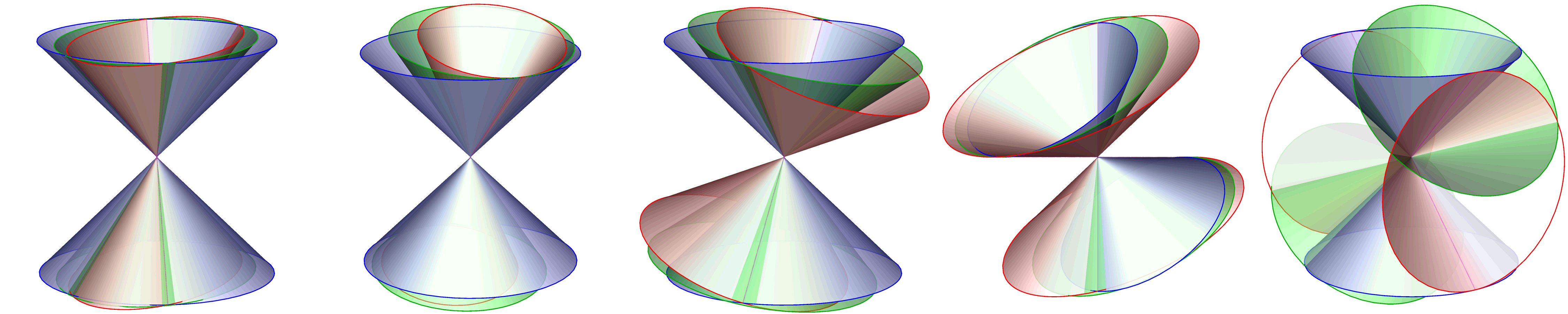}
      };
\end{tikzpicture}\vspace{-3mm}\protect\caption{\label{fig:jbs-1}The
  null cones of $h$, in green, relative to $f$ and $g$.}
\end{figure}

To see the geometrical meaning of the principal square root, consider
the covariant tensor,
\begin{align}
h_{\mu\nu} = g_{\mu\rho}\,(\sqrt{g^{-1}f}{)}^\rho{}_\nu\,,
\label{eq:h}
\end{align}
which is symmetric and nonsingular. For the principal square root,
$h$ is Lorentzian with the same signature as $g$ and $f$, and its null
cone contains the intersection of the null cones of $g$ and $f$. This
is depicted in Figure \ref{fig:jbs-1}, for Type I-IV metrics, with the
$h$ null cone shown in green. The explicit expressions are given in
Appendix \ref{app:proof}. The metric $h$ can be regarded as the
\textit{geometric mean} of $g$ and $f$, as shown in Appendix
\ref{app:geomMean}.\footnote{At the linearized level, $h$ is a
  combination of massive and massless modes \cite{HSvS1515}. But
  minimal coupling of matter to $h$ is not ghost-free since the vacuum
  energy contribution $\sqrt{|\det h|}=(|\det g/\det f|)^{1/4}$ is not
  found in the ghost-free bimetric potential.} As discussed above, $h$
is not a tensor in Type IV.

\begin{figure}[h]
\centering{}\hspace{2mm}\begin{tikzpicture}[x=1mm,y=1mm]
   \node[anchor=south west, inner sep=0] at (0,0) {
      \includegraphics[width=105mm]{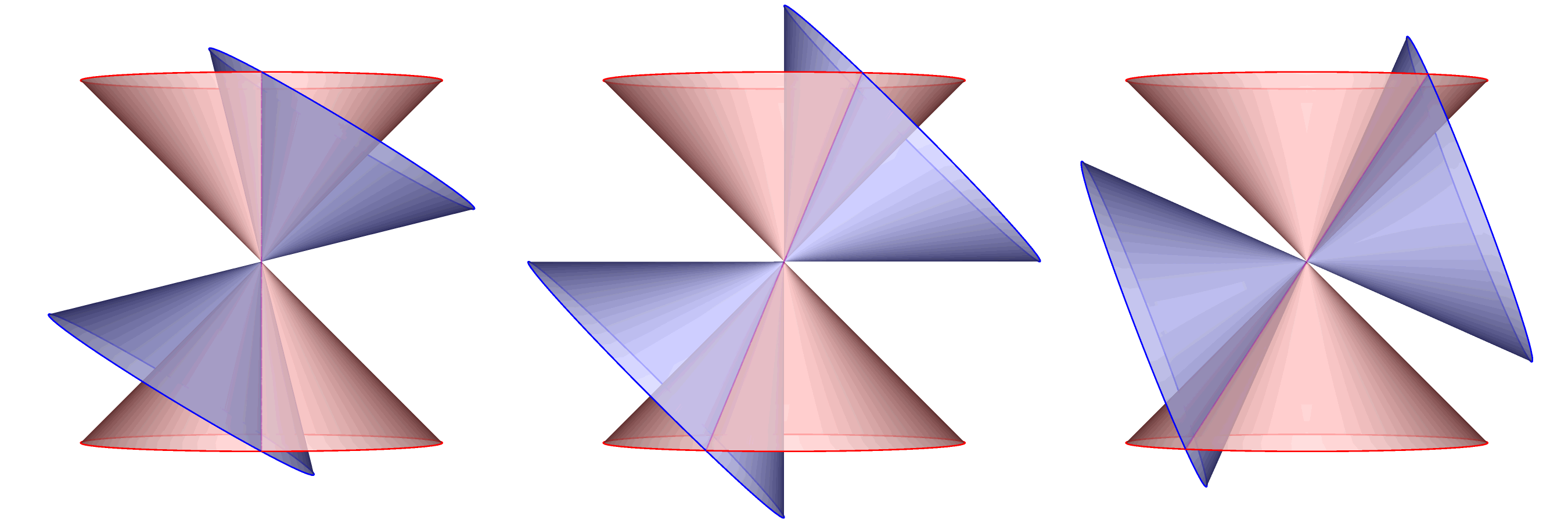}
      };
   \draw  [dashed,red] (0*35+3,17.54) -- (0*35+32,17.54);
   \draw  [dashed,red] (1*35+0,17.54) -- (1*35+35,17.54);
   \draw  [dashed,red] (2*35+3,17.54) -- (2*35+32,17.54);
   \node[] at (0*35+5,0) {(a)};
   \node[] at (1*35+5,0) {(b)};
   \node[] at (2*35+5,0) {(c)};
\end{tikzpicture}\vspace{-4mm}\protect\caption{\label{fig:bad-coords}
Good and  bad coordinate systems}
\end{figure}
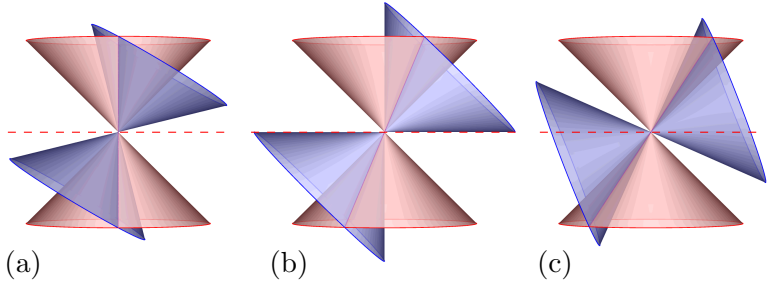

This property of $h_{\mu\nu}$ can be exploited to find coordinates
compatible with the 3+1 decompositions of both metrics. In general
relativity, as null cones tilt during evolution, one may need to
adjust the coordinate system to maintain a proper 3+1 decomposition,
for example, across a blackhole horizon. In bimetric theory similar
situations arise also for the relative tilt of the two null cones, as
illustrated in Figure \ref{fig:bad-coords}.
Let us start with a coordinate system where both metrics admit proper 3+1 
decompositions in Figure \ref{fig:bad-coords} (a),
i.e., $g^{00}<0$ and $f^{00}<0$. In Gaussian coordinates adapted
to $g$, a `horizontal' hypersurface (the dashed line) is initially
spacelike with respect to both metrics. As the $f$ null cone tilts and
hits the spacelike surface (b), its lapse
and shifts become infinite simultaneously. Beyond that, in (c), 
its lapse becomes imaginary ($f^{00}>0$), and
$f_{ij}$ becomes indefinite. This can be avoided by adapting the
coordinates to $h$ instead. Namely, a time direction at the `center' of the
$h$ null cone is always within the intersection of the $g$ and $f$
null cones, and a hypersurface transverse to it is always spacelike
with respect to all three metrics. One can implement this by starting
from the local coordinates of subsection \ref{sec:pair} which puts $f$
and $g$ in the canonical pair form, and then change to a new basis
where $h$ is, say, in Gaussian normal form.
In such a basis, both $f$ and $g$ will have a proper 3+1
decomposition.\footnote{It is instructive to revisit the ghost
  analysis of \cite{HRS1109} in the light of this discussion.
  There, initially one starts with a 3+1 type decomposition without
  specifying the sign of $f^{00}$. The decomposition is \textit{proper }
  only if $f^{00}<0$ so that the lapse $M=(-f^{00})^{-1/2}$ is
  real. Then a real square root exists if $x=-f^{00}-n^i f_{ij}n^j >0$
  for the given $n^i$ (see eqs.\,(3.22) and (3.30) in
  \cite{HRS1109}, also \cite{HKS}). However, $x>0$ holds not only
  for proper 3+1 decompositions but also when $f^{00}>0$, since then
  $f_{ij}$ is indefinite.  The present analysis shows that this case
  corresponds to a bad coordinate choice, and that at long as $x>0$, a
  frame always exists in which the decomposition is proper.}
Implementing this in massive gravity with a fixed $\fmn$ requires the
St\"{u}ckelberg fields.


\subsection{Type IV metrics as limits of Type IIb}
\label{sec:noIV}

Type IV configurations are not an isolated class in the algebraic
sense, but arise as limits of Type IIb metrics.  Here we point out
that when Type IIb metrics degenerate into Type IV, $S=\sqrt{g^{-1}f}$
develops a branch cut.  As will be argued in the next subsection, a
consequence is that Type IV metrics are not valid solutions in
bimetric theory.

Type IIb metrics arise when $g^{-1}f$ has complex eigenvalues,
$a\pm\ii b=\lambda(\cos\theta\pm\ii\sin\theta)$, with $b\neq 0$
or $\lambda >0\,, \theta\neq0,\pi$. Then, in a real Jordan basis, 
equation \eqref{type2b} gives,
\begin{align} 
g^{-1}f=\diag(\begin{pmatrixr} a & -b \\[-0.2em] b & a \end{pmatrixr},
\lambda_2,\lambda_3) =
\diag(\lambda\begin{pmatrixr} \cos\theta &\; -\!\sin\theta \\[-0.2em]
\sin\theta &\; \cos\theta\end{pmatrixr},\lambda_2,\lambda_3)\,. 
\label{eq:IIb2}
\end{align} 
For brevity, we again use $g$ for $Z^\tr g Z$, etc., since the $Z$
factors do not alter the outcome.  The nontrivial block of this is a
rotation matrix.  On the other hand, Type IV metrics \eqref{typeIV}
arise when $g^{-1}f$ has a pair of real negative eigenvalues,
\begin{align} 
g^{-1}f=\diag(-\lambda, -\lambda, \lambda_2,\lambda_3)\,,\qquad 
{\rm with}\quad \lambda,\lambda_2,\lambda_3>0. 
\label{eq:IV2}
\end{align} 
Clearly, Type IV can arise from Type IIb when $\theta\to\pi$, or
equivalently, $a<0$ and $b\to 0$ (the limit $\theta\to 0$ leading to   
Type I is harmless here).

\begin{figure}[h]
\centering{}\hspace{2mm}\begin{tikzpicture}[x=1mm,y=1mm]
   \node[anchor=south west, inner sep=0] at (0,0) {
      \includegraphics[width=90mm]{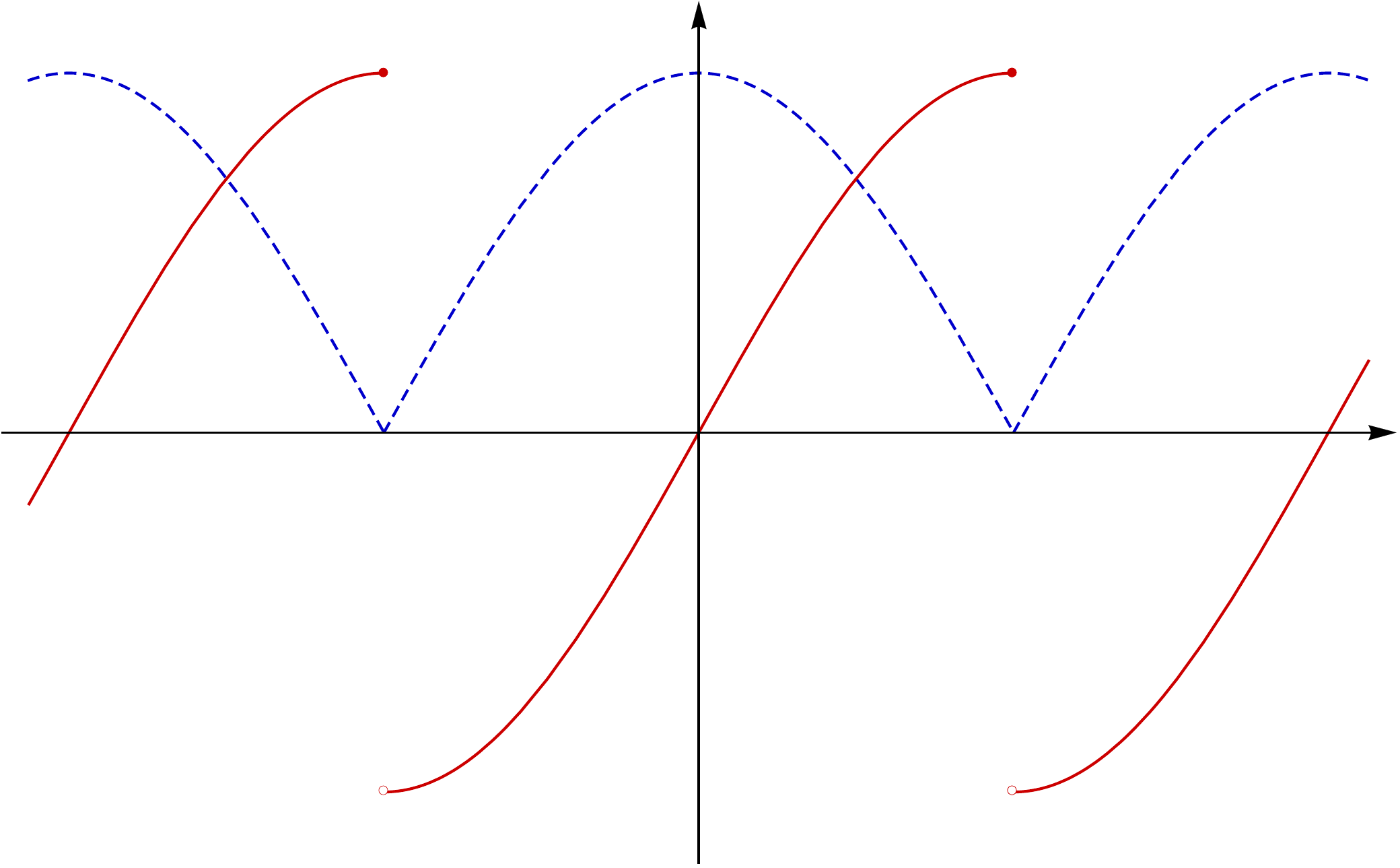}
      };
   \node at (42.8,54.2) {$S$};
   \node at (89,25) {$\theta$};
   \node at (24,25.5) {$-\pi$};
   \node at (47,25.5) {$0$};
   \node at (65.2,25.5) {$\pi$};
   \node [color=black!30!red] at (15,12) {$S^v_{~u}=\Im\sqrt{a+\ii b}$};
   \node [color=black!30!blue] at (94,45) {$S^u_{~u}=\Re\sqrt{a+\ii b}$};
\end{tikzpicture}\vspace{-3mm}\protect\caption{\label{fig:IIb-IV}
Type IV ($\theta=\pi$) as a branch cut of Type IIb.
}
\end{figure}
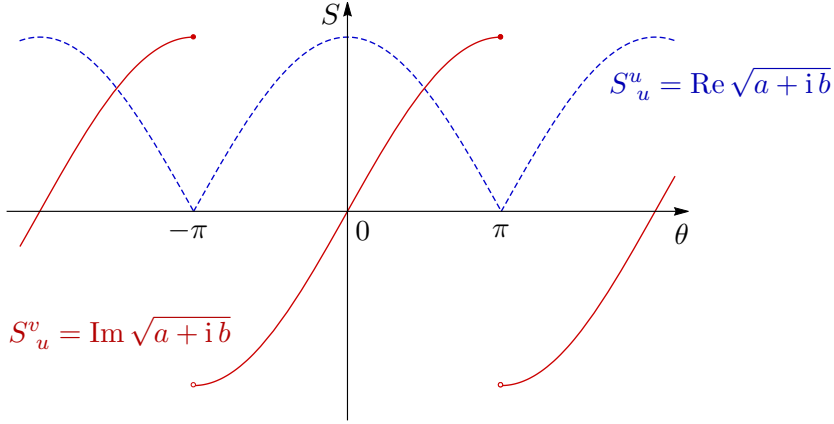

We are interested in $S=\sqrt{g^{-1}f}$ in this limit. Let us first
consider the square root $\sqrt{a+\ii b}\equiv \sqrt{\lambda}
\sqrt{\cos\theta +\ii\sin\theta}$. On the principal branch it has the
real and imaginary parts,\footnote{Recall that on the principal
  branch, the real part of $\sqrt{a+\ii b}$ is positive while its
  imaginary part has the same sign as $b$. Thus, the principal square
  root of $\ee^{\ii\theta}$ is 
  $\sgn(\cos\frac{\theta}{2}) \ee^{\ii\theta/2}$
  and not just $\ee^{\ii\theta/2}$.}
\begin{align} 
\Re\sqrt{a+\ii b} &
  =\frac{1}{\sqrt2}\sqrt{\sqrt{a^2+b^2} + a}
  =\sqrt{\lambda}\,\sgn(\cos\tfrac{\theta}{2})\cos\tfrac{\theta}{2}
  \,,\label{eq:Re-a+ib} 
\\
\Im\sqrt{a+\ii b} &
  = \frac{1}{\sqrt2}\sgn(b)\sqrt{\sqrt{a^2+b^2} - a}  
  =\sqrt{\lambda}\,\sgn(\cos\tfrac{\theta}{2})\sin\tfrac{\theta}{2}
\,.\label{eq:Im-a+ib}
\end{align} 
Here, $\sqrt{z}$ denotes the principal root of $z$ (another branch of
the square root is $-\sqrt{a+\ii b}$). These quantities are plotted in
Figure \ref{fig:IIb-IV}, for $0\leq \theta\leq 2\pi$.
The quantity $\Im\sqrt{a+\ii b}$ is discontinuous at $\theta=\pi$,
corresponding to the branch cut along the real negative axis on the
complex plane, whereas $\Re\sqrt{a+\ii b}$ is continuous but not
differentiable at the branch cut. By convention, the square root of
$-1$ is chosen as $+\ii$.

Now, the principal square root of $g^{-1}f$ for Type IIb
configurations \eqref{eq:IIb2} can be explicitly evaluated (see
also \cite{Higham08, Horn94}), and contains the 2$\times$2 block, 
\begin{align} 
\begin{pmatrixr} S^u_{~u} && S^u_{~v} \\[-0.2em] 
                    S^v_{~u} && S^v_{~v} \end{pmatrixr}
=\sqrt{\begin{pmatrixr} a && -b \\[-0.2em] b && a \end{pmatrixr}}
=\begin{pmatrixr} \Re\sqrt{a+\ii b} && -\Im\sqrt{a+\ii b}
\\[.1cm] \Im\sqrt{a+\ii b} && \Re\sqrt{a+\ii b} \end{pmatrixr}\!.
\label{eq:sqrtIIb}
\end{align} 
Here, $u$ and $v$ are the null cone coordinates $x^0\pm x^1$
corresponding to the form of $\gmn$ in \eqref{type2b}.  From Figure
\ref{fig:IIb-IV} it is clear that at $\theta=\pi$, when Type IIb goes
over to Type IV, the diagonal elements $S^u_{~u}, S^v_{~v}$ are not
smooth while the off-diagonal elements $S^u_{~v}, S^v_{~u}$ are
discontinuous. Thus, at $\theta=\pi$ (Type IV), the variation
of the square root matrix $S$ with respect to $\theta$ is not
defined. The same holds for the bimetric potential $V(S)$.


\subsection{The absence of null coupled (Type IV) configurations}  

So far we have argued that Type IV configurations could be discarded
on the ground that they lead to the violation of general covariance. This
justifies not using them as initial conditions, or as ansatz for
finding solutions.  However, since Type IV metrics can also arise as
limits of Type IIb, they can be discarded consistently only if the
equations of motion do not evolve Type IIb metrics into Type IV.  Here
we show that this is indeed the case and that the theory does not
admit null coupled (Type IV) configurations as solutions due to the
square root branch cut.  This is illustrated with a simple example
from mechanics.

The variation of the bimetric action \eqref{BiM} with respect to
continuous functions $\gmn$ has the form $\delta_g{\cal S}=\int d^4x
(G^g_{\mu\nu}+V^g_{\mu\nu})\delta g^{\mu\nu}$, where $G^g_{\mu\nu}$ is
the Einstein tensor and $V^g_{\mu\nu}$ arises from the variation of
the potential $\sqrt{-g} V$. Provided that $V^g_{\mu\nu} $ are also
continuous functions, the fundamental lemma of calculus of variations
states that requiring $\delta_g{\cal S}=0$ leads to the equations of
motion (similarly for $\fmn$),
\begin{gather}
G_{\mu\nu}^{g} + V_{\mu\nu}^{g} =0\,,\qquad
G_{\mu\nu}^{f} + V_{\mu\nu}^{f}=0\,.
\label{eq:bim-eom}
\end{gather}
Let us assume that these admit solutions $g$ and $f$ that evolve into
Type IV configurations in some region $\Sigma$ of spacetime. But in
this region, the potential $V$ could not have been varied to produce
the $V^g_{\mu\nu}$ ($V^f_{\mu\nu}$) in $\delta_g{\cal S}$
($\delta_f{\cal S}$), hence the above equations do not apply in
$\Sigma$. This is a consequence of $S=\sqrt{g^{-1}f}$ being the
principal square root where Type IV coincides with a branch cut.

In the parameterization of the previous subsection, the branch cut is
at $\theta(x)=\pi$ which, generically, corresponds to a 3 dimensional
hypersurface $\Sigma$ characterized by the normal $\p_\mu\theta(x)$.
Since the variation of $\sqrt{-g}\,V(S)$ is discontinuous at
$\theta(x)=\pi$, the variational principle can be applied piecewise
for $\theta(x)<\pi$ and $\theta(x)>\pi$, or formally, on both sides of
the hypersurface defined by $\theta(x)=\pi$, to yield two sets of
equations of motion,
\begin{align}
&\theta<\pi :\qquad G_{\mu\nu}^{g} + V_{\mu\nu}^{g}(S_<) =0\,,\quad
G_{\mu\nu}^{f} + V_{\mu\nu}^{f}(S_<)=0\,, 
\\[.1cm]
&\theta>\pi :\qquad  G_{\mu\nu}^{g} + V_{\mu\nu}^{g}(S_>) =0\,,\quad
G_{\mu\nu}^{f} + V_{\mu\nu}^{f}(S_>)=0\,,
\label{eq:bim-eom2}
\end{align}
where $S_{<}$ and $S_{>}$ are $S$ evaluated on two sides of the branch
cut. But there are no equation valid at $\theta=\pi$.  Note that the
Bianchi constraints $\nabla^{g}_\mu V^{g\mu}{}_\nu=0$ and
$\nabla^{f}_\mu V^{f\mu }{}_\nu=0$ are also not valid at $\theta=\pi$.
The variational principle does not lead to equations of motion valid
for Type IV metrics showing that such configurations are not valid
solutions of bimetric equations.

If a Type IIb solution, say, for $\theta<\pi$, approaches Type IV, it
must either be reflected back at the branch cut, or must be matched
across $\theta=\pi$ to another Type IIb configuration for
$\theta>\pi$, in the spirit of junction conditions with no need for
equations valid at $\theta=\pi$. Which of the two scenarios is
actually realized is not addressed here. Matching Type IIb
configurations across Type IV could potentially lead to a problem
with the relative time orientation of the two metrics, but 
restricting $S$ to a single branch also preserves the relative time  
orientations (as discussed in Appendix \ref{sec:a-Orientability}). 

Here we do not attempt to analyze the behavior of bimetric solutions
near the branch cut in more detail, but it is instructive to
illustrate the situation with an example. A simple mechanical system
displaying a similar behavior is given by the action,
\begin{align}
A=\int \dd t\left(\dot{x}^2/2 -\lambda \sqrt{x^2}\right)\,, 
\end{align}
defined on the principal branch of $\sqrt{x^2}=|x|$. This describes a
charge in a constant electric field along the $x$-axis which abruptly
reverses direction at $x=0$. The variation of the action is
discontinuous at $x=0$ and the variational principle yields the
equation of motion,
\begin{align}
\ddot{x}=-\lambda\quad (x>0)\,,\qquad 
\ddot{x}=\lambda\quad (x<0), 
\label{eq:example}
\end{align}
with no equation valid at $x=0$. However, continuous solutions can be
easily obtained by matching the two parabolic solutions at $x=0$. For
$\lambda<0$, the solution crosses $x=0$ at most once. For $\lambda>0$,
the solution $x(t)$ is an array of alternating concave and convex
parabolas matched at $x=0$, describing a particle oscillating about
$x=0$.

In the mechanical system \eqref{eq:example}, the discontinuity at
$x=0$ is to be regarded as the outcome of a coarse graining
approximation. At the microscopic level, the electric field of
strength $\lambda$ cannot be reversed abruptly but will change
direction over a thin layer around $x=0$. Equations
\eqref{eq:example} then provide an effective description at large
distances. In analogy, the appearance of Type IV discontinuity may be
regarded as an indication of the effective nature of bimetric theory,
signaling the need for new degrees of freedom at a more fundamental
level.\footnote{We would like to thank K. Izumi and Y. C. Ong for a
  discussion of this issue.}

Finally, it is worth noting that in the analysis of perturbations in
massive gravity \cite{Bernard:2015mkk}, one needs to solve the matrix
equation $\delta(S^2)=S\,\delta S +\delta S\,S$ to express $\delta S$
in terms of $\delta(S^2)$. By a theorem of Sylvester, a unique
solution exists if and only if $\sigma(S) \cap \sigma(-S) =
\varnothing$, i.e., when $S$ and $-S$ have no common eigenvalues.
This property is not satisfied by Type IV metrics, which were then
excluded from the analysis.


\section{Remarks on causality and superluminality}
\label{sec:causality}

To put the results in perspective, we briefly comment on the problems
of causality and superluminality in bimetric theory, and discuss the
implications for some studies of acausality in massive gravity
\cite{IO, DIOWredux,DIOW1312,DSWZ1408,DIOWproblems}. For convenience,
the condition for bimetric theory to have meaningful dynamics can be
stated in two steps.
\begin{itemize}
\item The covariant equations \eqref{eq:bim-eom} must be writable as
  time evolution equations with spatial hypersurfaces for specifying
  initial data, at least locally. As discussed earlier, this is not
  self-evident in a bimetric setup.
\item The initial data must uniquely determine the solutions at later
  times, thereby implying causality.
\end{itemize}
We have addressed the first requirement, showing that the theory
admits 3+1 decompositions compatible with both metrics. This provides
the necessary setup for addressing the second condition, but the
well-posedness of the initial value problem is not addressed here and
remains an open question. However, one can already note aspects of
the bimetric theory that differ from general relativity, but may not
necessarily imply lack of causality.

If the initial conditions (specified over an appropriate domain of
dependence) uniquely determine the solution at later times then the
dynamics will be causal even if the theory contains ``faster than
light'' propagation, as explained in detail in \cite{Geroch}. In such
cases, the equations will determine a causal cone which need not
coincide with the null cone of a gravitational metric, say,
$\gmn$. Then, superluminality with respect to $\gmn$ will not
necessarily indicate causality violation.  In traditional single
metric theories, including in general relativity, the equations of
motion encode a causal cone that coincides with the null cone of
$\gmn$ which thus determines the causal structure. However, as
emphasized in \cite{Geroch}, this need not be the case in more general
setups.  In \cite{Babichev:2007dw} the possibility of superluminal
propagation without causality violation has been discussed in a class
of scalar field theories. For discussions of similar issues in some
other bimetric scenarios see \cite{Schuller:2016onj,Drummond:2013ida}.

In the ghost-free bimetric theory, one expects that the
Einstein-Hilbert term for each metric allows for propagation at least
within the null cone of that metric.  This alone means that, whenever
the two null cones do not coincide, the theory could allow for the
propagation of, say, $f$-metric perturbations or matter, outside the
null cone of the $g$-metric, and \textit{vice versa}. Interactions
introduce further complexities, but spacelike propagation with respect
to either or both metrics is still expected. However, as indicated above,
this is not necessarily a violation of causality, provided the initial
value problem is well-posed. This problem has not yet been analyzed in
a conclusive way for bimetric theory. For most practical purposes,
where the theory should be close to GR, the effects of propagation
outside the null cone of the gravitational metric can be suppressed.
Also, since observable matter couples minimally only to a single
metric, say, $\gmn$, in a ghost-free way, it is primarily sensitive to
the null cone of this gravitational metric.  For cosmologically
relevant solutions, the $g$ and $f$ null cones approach each other at
late times.

In any case, it should be emphasized that the bimetric potential is
the spin-2 analogue of the Proca mass for vector fields,
$\sqrt{-g}g^{\mu\nu}A_\mu A_\nu$, indicating that the theory may be
incomplete in the absence of some extra Higgs-like degrees of
freedom. It is possible that, in the spin-2 case, ignoring the extra
degrees of freedom could cause pathologies even at the classical
level. One may also have to further restrict the allowed solutions by
imposing the analog of energy and hyperbolicity conditions, as in
general relativity.

\subsection{On the studies of acausality in massive gravity}

Here, after a brief review of massive gravity in the vielbein
formulation and of its initial value formulation, we point out that a
number of arguments for acausal propagation in the theory is based on
the Type IV and other inadmissible configurations.  Thus more work,
perhaps on the lines of \cite{IO}, is needed to clarify the causal
properties of the theory (see also \cite{Volkov1,Volkov2}).

Massive gravity theories can arise as limits of bimetric theory ($m_f
\rightarrow \infty$ at fixed $m_g$) around classes of solutions that
admit such a limit \cite{HSvS1407}. In practice, one simply drops
$\sqrt{|f|}R_f$ from the bimetric action \eqref{BiM} and treats
$f=\bar f$ as nondynamical (e.g., $\bar f=\eta$, \cite{dRGT} or
any fixed $\bar f$, \cite{HR1103,HRS1109}).\footnote{If massive
  gravity is to have $g\propto \bar f$ as a solution, then $\bar f$
  must solve Einstein's equation and only then, perturbations $\delta
  \gmn$ will have a Fierz-Pauli mass. An Einstein equation for $\bar
  f$ arises in the massive gravity limit of bimetric equations, but
  not from the massive gravity action.}
In terms of vielbeins $e^a_\mu$ and $d^a_\mu$, corresponding to
metrics $\gmn$ and $\fmn$, the interaction potential becomes \cite{HR,
  Zumino},
\begin{align}
   \sum_{n=0}^4 \beta_n \; 
   \epsilon_{a_1\cdots a_{n} a_{n+1} \cdots a_{4}} \,
   \epsilon^{\mu_1\cdots \mu_n \mu_{n+1} \cdots \mu_{4}} \;
   d^{a_1}_{\mu_1} \cdots d^{a_n}_{\mu_n} \,
   e^{a_{n+1}}_{\mu_{n+1}}\cdots e^{a_4}_{\mu_4}\,.
\end{align}
A symmetrization constraint, $d^{~}_{\mu {[a}} e^\mu_{b]}=0$, implies
$\sqrt{g^{-1}f}=e^{-1}d$, insuring equivalence to the metric
formulation. Although the vielbein form seems simpler to manipulate, 
in reality, solving the symmetrization constraint involves the same
level of complication as computing a square root matrix.
When $f$ is nondynamical, there is no \textit{a priori} requirement for
using compatible 3+1 decompositions for $f$ and $g$.  However, such
decompositions always exist and must be employed when the theory is
regarded as a bimetric limit.  Below we comment on some analysis of
the initial value problem, superluminality, and acausality in massive
gravity, carried out mostly in the vielbein formulation.

The \textit{initial value problem} in massive gravity was addressed in
\cite{IO}, by selecting a time direction and expressing the equations
in the quasilinear form $A_{ab}\p_t\phi^b=F_a$, where $A_{ab}$ and
$F_a$ are functions of the dynamical fields $\phi^a$ and their spatial
derivatives. The well-posedness of the initial value problem then
requires that the matrix $A$ is invertible, so that all $\phi^a$ have
well-defined evolution equations. However, if the theory admits field
configurations such that $\det A=0$, then time evolution will not be
well-defined. Further analysis can shed light on whether valid acausal
configurations can arise in massive gravity, or if further
restrictions are needed. In any case, the analysis of \cite{IO} shows
that configurations with $\det A=0$ are not a generic feature of
the theory, as had been suggested earlier.

\textit{Superluminality} and \textit{acausality} in massive gravity have
been studied in reference \cite{DIOWredux} and related works by
searching for characteristic surfaces (corresponding to maximum speed
signals) that are spacelike with respect to $\gmn$. Propagation along
such a surface would be superluminal with respect to $\gmn$. Although
this, by itself, does not imply acausality, \cite{DIOWredux} argues
that specific examples can be constructed where superluminality leads
to the formation of local Closed Causal Curves (CCC). Here, we show
that these arguments rely on disallowed field configurations
corresponding to the nonprincipal square roots.
 
Spacelike characteristic surfaces (with timelike normals) exist if an
associated characteristic matrix, say, $\Delta$, obtained from the
equations of motion, satisfies $\det\Delta=0$.\footnote{In general,
  the well-posedness of the initial value problem requires the
  vanishing of the characteristic determinant. This also determines
  the causal cone for the problem, see, for example,
  \cite{Bojowald}. The analysis of characteristics in
  \cite{DIOWredux} has a more limited scope.}  Reference
\cite{DIOWredux} contains two examples of field configurations that are
supposed to give $\det\Delta=0$. In the first example, this condition
reduces to $-3m^2/2+l_{00}K^j K_j=0$, for some vectors $K^j$. It is
stated that if the normal to the characteristic surface (which has
been taken to be timelike with respect to $\gmn$) is not timelike with
respect to $\fmn$, then $l_{00}$ is positive. Then the equation
becomes a difference of positive terms and can be satisfied; hence
the superluminality. However, this scenario is not realized by the
allowed configurations. In our conventions,
$l_{\mu\nu}=(gh^{-1}g)_{\mu\nu}$, where $h=gS$ is the metric in
\eqref{eq:h}, depicted in Figure \ref{fig:jbs-1}.  As is obvious from
the figures (and can be explicitly verified from the equations in full
generality), any coordinate system in which $g$ and $f$ admit
compatible 3+1 decompositions leads to $l_{00}<0$.  For example, in
Gaussian normal coordinates adapted to $g$,
$l_{00}=h^{00}=-N_h^{-2}<0$, where $N_h$ denotes the lapse of $h$.
Obviously, $l_{00}=h^{00}<0$ as long as the spatial hypersurface is
spacelike with respect to $h_{\mu\nu}$ (even for a time direction
outside the null cone of $h$). As the $h$ null cone tilts away from
the $g$ null cone, the $f$ null cone will hit the spatial hypersurface
(invalidating the coordinate system choice) before $l_{00}$ turns
positive. Hence, in this example, the superluminality constraint is not
satisfied.  
 
The second example in \cite{DIOWredux} is about the formation of
Closed Causal Curves (CCC), which would imply violation of local
causality. This is elaborated further in \cite{DIOW1312}. However,
this construction requires that $g$ and $f$ are Type IV metrics which,
as shown above in Section \ref{sec:implications}, are not valid
solutions of the theory.

Reference \cite{DIOWproblems} provides new examples of backgrounds
that exhibit superluminality and Closed Causal Curves, and which are
not Type IV configurations. However, these correspond to choosing a
nonprincipal square root which is ruled out if massive gravity is
regarded as a limit of bimetric theory (but could be allowed if no
extra symmetries are required). These examples amount to choosing the
vielbeins as $(e^a_{~\mu}) =\diag( A, B, C, -B)$ and
$(d^a_{~\mu})=\diag(1,1,1,1)$.  Note that the metric $g=e^\tr\eta e$
does not distinguish between vielbein $e$ above and some $e'=\diag(
A, B, C, B)$ which result in opposite volume orientations
$\det(e)=-\det(e')$. Also in both cases, $S^2=g^{-1}f=\diag(A^{-2},
B^{-2}, C^{-2}, B^{-2})$.  But the square root matrix $S$ is
sensitive to the difference; $S=e'^{-1}d$ is a primary square root,
whereas $S=e^{-1}d$ is a nonprimary one which we have ruled
out.\footnote{The statement in \cite{DIOWproblems}, that a bimetric
  theory cannot be consistently coupled to matter without destroying
  its degree of freedom count, is inaccurate since ghost-free matter
  couplings are well known \cite{HR1109}.}

Hence, to conclude this section, the above references do not prove
that massive gravity violates causality through the formation of local
CCCs, especially, when the theory is regarded as a limit of bimetric
theory and defined as outlined above. Settling the question of
causality in bimetric theory requires more accurate analysis that
takes into account the validity of the field configurations, and of
the 3+1 decompositions. For alternative approaches see, for example,
\cite{Camanho:2016opx}.

\section{Discussion}
\label{sec:discussion}

Here we briefly highlight some ramifications and limitations of our
results.

The above results show the validity of the proofs of the absence of
the BD ghosts in massive gravity and bimetric theory. The proofs were
carried out in the Hamiltonian formulation assuming that the two
metrics admitted simultaneous 3+1 decompositions
\cite{HR1106,HRS1109,HR1109,HR1111,Alexandrov} (for a detailed
analysis see \cite{HL}). The analysis here shows that this is fully
adequate since all allowed configurations admit such a
decomposition. The ghost analysis has also been carried out in
covariant forms for massive gravity \cite{cedrik,DSWZ1408} and,
bimetric theory \cite{Bernard:2015uic}. In these analyses too, except
for massive gravity models with $\beta_3=0$, one needs to invoke 3+1
decomposition to verify the validity of the constraints.

Early studies of ghost in massive gravity involved working with a
binomial series expansion of $\sqrt{\II+h}$ that holds for
configurations perturbatively close to $\gmn=\eta_{\mu\nu}$
\cite{dRGT}. As long as $h$ remains perturbative, this is consistent
with restricting the square root to the principal branch (since $1-x$
and $|1-x|$ are the same for $x <1$ but not for $x>1$). However, if
the square root is defined without restricting to a specific branch
(so that $\sqrt{\ee^{\ii\theta}}=\ee^{\ii\theta/2}$, rather than
$\sqrt{\ee^{\ii\theta}}=|\ee^{\ii\theta/2}|$), then its variation is
continuous and Type IV metrics become regular solutions of the field
equations. Restricting the square root to the principal branch avoids
this problem. The choice of the branch for the square root matrix may
also be restricted by ghost considerations and by requiring
backgrounds with certain symmetries as in \cite{Comelli:2015ksa} for
the case of Lorentz invariant massive gravity.

The results also clarify the meaning of the vielbein symmetrization
condition $d^{~}_{\mu {[a}} e^\mu_{b]}=0$ \cite{Zumino}, sometimes
called the Deser-van Nieuwenhuizen gauge, although it is known not to
be a gauge condition. Since this condition is equivalent to finding
the matrix square root $S$ \cite{Deffayet:2012zc}, it involves the
same challenges of multiple solutions and the choice of branch, with
the same resolutions as proposed here. Our results show that the
symmetrization condition can be regarded as a gauge condition, but
only when the two vielbeins admit simultaneous 3+1
decompositions. This also provides a physical justification for
imposing the symmetrization condition by hand in theories where they
are not necessarily implied by the equations of motion, such as in the
vielbein formulation of bimetric theory with all $\beta_n\neq 0$,
\cite{HR,HSvS1204} or in \cite{AH1609}.

The presented Theorem \ref{the:interlock}, which connects the algebraic
existence of the square root $\sqrt{g^{-1}f}$ with the local
geometrical relations between the null cones of $f$ and $g$, is
actually more general. In fact, the proof applies to any function
$F(g^{-1}f)$ which requires that the eigenvalues of $g^{-1}f$ are not
on $\mathbb{R}^{-}$. For example, $F(X)$ can be taken as the function
$X^{\xi}\equiv\exp(\xi\log X)$ for a real $\xi$. 

As emphasized, the statements in this paper are local and can serve as
a first step towards studying causality and the initial value problem
at the global level. Since global hyperbolicity is a reasonable
condition in the context of GR, the logical extension would be to
define a notion of simultaneous global hyperbolicity for a manifold
equipped with two metrics and consistent with the bimetric equations
of motion. Causality in massive gravity has recently been considered
in \cite{Camanho:2016opx} from a different point of view.

Also, as pointed out, the appearance of the branch cut associated with
Type IV (and the analogy with a discontinuous electric field) could be
an indication of the effective nature of the theory and could shed
light on the nature of the extra degrees of freedom that have been
integrated out. Then one expects that this may also have ramifications
for the specific form of the quantum completion of the theory
\cite{Cheung:2016yqr}. 


\acknowledgments

We would like to thank Keisuke Izumi, Yen Chin Ong, Angnis
Schmidt-May, Kjell Rosquist, Lars Andersson, Anders Lundkvist, and
Luis Apolo for helpful discussions. We are grateful to Mikael von
Strauss for a careful reading of the manuscript. We would also like to
thank the organizers of the PI conference on ``Superluminality in
Effective Field Theories for Cosmology'' 2015, where the results in
Section 2 were first presented.  Subsequently, the results have been
presented in other workshops and conferneces, including FANCY 2016 in
Odense, Indian Strings Meeting 2016 in Pune, and a number of Nordita
workshops.

\clearpage
\phantomsection
\addcontentsline{toc}{section}{Appendices}
\part*{\Large Appendices}

\appendix

\def\toclevel@section{1} 
\def\toclevel@subsection{2}
\addtocontents{toc}{\string\let\string\l@section\string\l@subsection}


\section{The square root matrix function}
\label{app:f-struct}

Here we collect some properties of the square root matrix function
that are used in the text of the paper (see, for example,
\cite{Higham08}). Given an ordinary scalar function $F(x)$ and a
matrix $A$, one can construct a matrix function $F(A)$. An example is
$F(x)=\sqrt{x}\,$, leading to the matrix square root,
$F(A)=\sqrt{A}$. To make this precise, consider transforming $A$ to
its Jordan normal form, $Z^{-1}AZ=\diag(J_{1},\dots,\dots,J_{s})$,
where $J_{i}\equiv J_i(\lambda_i)$ is the Jordan block corresponding
to the eigenvalue $\lambda_i$ which could be real or complex. Then,
the precise meaning of $F(A)$ is a matrix which, in the same Jordan
basis, reads (Definitions 1.2 in \cite{Higham08} and 6.2.4 in
\cite{Horn94}),
\begin{align}
Z^{-1}F(A)Z &
=\diag(F(J_{1}),\dots,\dots,F(J_{p}))\,,
\label{eq:MatFunc}
\end{align}
where,\vspace{-0.2em}
\begin{equation}
F(J_{k})\equiv
\begin{pmatrix}F(\lambda_{k}) & F^{\prime}(\lambda_{k}) & \cdots & 
\frac{1}{(n_{k}-1)!}F^{(n_{k}-1)}(\lambda_{k})\\[-0.2em]
 & F(\lambda_{k}) & \ddots & \vdots\\[-0.2em]
 &  & \ddots & F^{\prime}(\lambda_{k})\\
 &  &  & F(\lambda_{k})
\end{pmatrix}\!.
\label{eq:SqrtFunc}
\end{equation}
Here $F(\lambda_{k})$ is the scalar function and $F^{(n)}$ denotes its
$n$-th derivative. When the stem function $F(x)$ has multiple branches,
the same branch must be chosen within each block.

For $F(x)=x^{1/2}$, equation \eqref{eq:MatFunc} defines the matrix
function $F(A)=\sqrt{A}$, with, 
\begin{equation}s
F(J_{k})=\pm\begin{pmatrix}
\lambda_{k}^{1/2} && \frac{1}{2}\lambda_{k}^{-1/2} & \cdots & 
(-1)^{n_k} \frac{(2n_k-5)!!}{2^{n_k-1}} \lambda_k^{3/2-n_k} \\
 && \lambda_{k}^{1/2} & \ddots & \vdots\\
 &&  & \ddots & \frac{1}{2}\lambda_{k}^{-1/2} \\
 &&  &  & \lambda_{k}^{1/2}
\end{pmatrix}\!.
\label{eq:SqrtJ}
\end{equation}
The $\pm$ signs in front of the matrix reflect explicitly the possible
choices of branch for $\lambda_i^{1/2}$ which must be the same within
a given block. For a real matrix $A$, its complex eigenvalues come in
conjugate pairs and, by Theorem 3.4.5 in \cite{Horn90}, the
corresponding Jordan blocks $J_k(\lambda), J_k(\lambda^*)$ can be
written in a real form $C_k(a,b)$ in a real basis $Z$, as in equation
\eqref{eq:cpt-3}. The conditions for the reality of $\sqrt{A}$ are
stated in subsection \ref{sec:realSQR}.

\textit{Primary, nonprimary and principal square roots:} A primary square
root is specified by the condition that, if the eigenvalues appearing
in two or more Jordan blocks of $A$ are equal, then in $F(A)$ the same
branch must be chosen for all such blocks. For the remaining blocks
with unequal eigenvalues, the choice of branch is unrestricted leading
to multiple primary square roots. All these are expressible as
polynomials in the matrix $A$ and have useful properties, in
particular, under a similarity transformation $Q$, they satisfy,
\begin{align}
F(Q^{-1} A Q) = Q^{-1} F(A) Q\,.
\label{eq:simPrimSqrt}
\end{align}   
For other functions $F(x)$ with multiple branches, the corresponding 
\textit{primary} matrix functions $F(A)$ are defined in a similar way.

If the matrix $A$ has no eigenvalues on the negative real axis
$\mathbb{R^-}$, then among its primary square roots there is a unique
one with all eigenvalues in the open right half-plane, corresponding
to choosing the principal branch for all blocks $F(J_k)$. This is the
\textit{principal} square root of $A$ which is real when $A$ is real. 

\textit{Nonprimary} square roots arise when $A$ has equal eigenvalues
that appear in different Jordan blocks, and we do not choose the same
branch for the corresponding $F(J_k)$. Then $A$ will have an infinite
number of nonprimary square roots. An important feature of nonprimary
matrix functions is that do not obey equation \eqref{eq:simPrimSqrt}.

If a matrix $A$ has negative eigenvalues, then it cannot have any real
primary square root, though it could have nonreal primary or real
nonprimary real roots.
Furthermore, a principal square root cannot be nonprimary.

A simple instructive example is the 2$\times$2 identity matrix
$A=I_2$ with two equal eigenvalues $+1$. The primary square
roots are $\pm I_2$ and satisfy \eqref{eq:simPrimSqrt}, with
$+I_2$ being the principal square root. A nonprimary square root
is $\diag(+1, -1)$ which clearly violates \eqref{eq:simPrimSqrt}. In fact,
$Q^{-1} \diag(+1, -1) Q$ generates an infinite number of
nonprimary square roots of $I_2$.


\section{Types of blocks in the theorem}
\label{app:proof}

In this section,
we scrutinize all the allowed configurations and explicitly 
construct a common timelike vector and a common spacelike surface
element relative to both metrics (exhibiting the causally coupled case),
or alternatively show that such construction is not possible
(resulting in two common null hyperplanes, i.e., exhibiting the null
coupled case). This finalizes the sufficiency proof of the theorem.
During this process, we also calculate the explicit form of a congruence
that diagonalizes the metric of the geometric mean, $h = gS$
(see Appendix \ref{app:geomMean}).

Let us denote $Z^{\tr}fZ$, $Z^{\tr}gZ$ and $Z^{\tr}hZ$ simply as
$f$, $g$ and $h$. The assumed parameter space is $\lambda,\lambda_{k}>0$,
$a\in\mathbb{R}$, $b\ne0\in\mathbb{R}$. The components of vectors in 
a tangent space (of which the null cones are subsets) will be denoted by 
$t\equiv x^{0}$, $x\equiv x^{1}$ and $y\equiv x^{2}$. In the following, 
the quadratic form $h=gS$ will be diagonalized by an additional congruence 
$X$, so that the common timelike vector for $f$, $g$ and $h$ is always set
to $\tau=X\cdot(1,0,0)^{\tr}$ while the common spacelike surface element 
to $\Sigma=X\cdot(0,x,y)^{\tr}$.

\paragraph{Type I.}

This is the case where $g^{-1}f$ could be fully diagonalized, and,
\begin{equation}
g=\diag(-1,1,1),\quad
f=\diag(-\lambda_{0},\lambda_{1},\lambda_{2}),\quad 
h=\diag(-\lambda_{0}^{1/2},\lambda_{1}^{1/2},\lambda_{2}^{1/2}).
\end{equation}
Here, the null cones of $f$ and $g$ are centered (without the shifts)
and may or may not intersect (depending on the combinations of 
$\lambda_{k}\gtrless1$).
Obviously, a congruence $X$ which diagonalizes $h$ is identity, and,
\begin{alignat}{2}
\tau^{\tr}g\tau & =-1<0, & \qquad\Sigma^{\tr}g\Sigma & =x^{2}+y^{2}>0,\\
\tau^{\tr}f\tau & =-\lambda_{0}<0, & \Sigma^{\tr}f\Sigma & =\lambda_{1}x^{2}+\lambda_{2}y^{2}>0,\\
\tau^{\tr}h\tau & =-\lambda_{0}^{1/2}<0, & \Sigma^{\tr}f\Sigma & 
=\lambda_{1}^{1/2}x^{2}+\lambda_{2}^{1/2}y^{2}>0.
\end{alignat}

\paragraph{Type IIa.}

The diagonalization of $g^{-1}f$ is suppressed by the presence
of a common tangent plane for the null cones of $f$ and $g$. In
this case, $g$ is given in null coordinates,
\begin{equation}
g=\diag(\epsilon\!\begin{pmatrix}0 & 1\\[-0.2em]
1 & 0 \end{pmatrix}\!,1),\ 
f=\diag(\epsilon\!\begin{pmatrix}0 & \lambda\\[-0.2em]
\lambda & 1
\end{pmatrix}\!,\lambda_{2}),\ 
h=\diag(\epsilon\!\begin{pmatrix}0 & \lambda^{1/2}\\[-0.2em]
\lambda^{1/2} & \,\frac{1}{2}\lambda^{-1/2}
\end{pmatrix}\!,\lambda_{2}^{1/2})\,.
\end{equation}
where $\epsilon=\pm1$. A congruence $X$ which diagonalizes $h$ reads,
\begin{equation}
X = \frac{1}{\sqrt2}\begin{pmatrix}
-\epsilon\,(1+\epsilon p)^{1/2} &\,& \epsilon\,(1-\epsilon p)^{1/2} \\
 (1-\epsilon p)^{1/2} & & (1+\epsilon p)^{1/2} \\
\end{pmatrix}\!,
\qquad
p \equiv (1+16\lambda^2)^{-1/2}\,,
\end{equation}
where $0<p<1$ for $\lambda>0$.
Consequently,
\begin{alignat}{2}
\tau^{\tr}g\tau & = -(1-p^2)^{1/2} < 0, & 
\qquad\Sigma^{\tr}g\Sigma & = (1-p^2)^{1/2} x^2+y^2 > 0,\\
\tau^{\tr}f\tau & = -\frac{(1-\epsilon p)^2}{4p} <0, & 
\Sigma^{\tr}f\Sigma & = \frac{(1+\epsilon p)^2}{4p} x^2+\lambda_2 y^2>0,\\
\tau^{\tr}h\tau & =
-\frac{ 1- \epsilon p}{2 p^{1/2} (1-p^2)^{1/4}} 
< 0, & 
\Sigma^{\tr}h\Sigma & =
\frac{ 1+ \epsilon p}{2 p^{1/2} (1-p^2)^{1/4}} 
x^2+\lambda_2^{1/2}y^2>0.
\end{alignat}

\paragraph{Type IIb.}

The null cone of $f$ is rotated and dilated by $a,b$, where we can
parameterize $a=\lambda\cos\theta$, $b=\lambda\sin\theta$ with $\lambda=\sqrt{a^2+b^2}>0$
and $\theta\ne0\in(-\ppi,\ppi)$.
(For Type IIb, necessarily $b\ne0$, otherwise we have Type I or Type IV.)
Then,
\begin{gather}
g=\diag(\begin{pmatrix}0 & 1\\[-0.2em]
1 & 0
\end{pmatrix}\!,1),\quad f=\diag(\begin{pmatrix}\lambda\sin\theta & \,\lambda\cos\theta\\
\lambda\cos\theta & \,-\lambda\sin\theta
\end{pmatrix}\!,\lambda_{2}),\\[0.3em]
h = \diag(\lambda^{1/2}\begin{pmatrix}\sgn(\cos\varphi)\sin\varphi & \,\lvert\cos\varphi\rvert\\
\lvert\cos\varphi\rvert & \,-\sgn(\cos\varphi)\sin\varphi
\end{pmatrix}\!,\lambda_{2}^{1/2}),\quad \varphi\equiv\frac{\theta}{2}.
\end{gather}
A congruence $X$ which diagonalizes $h$ reads,
\begin{equation}
X=\begin{pmatrix}
-\frac{1}{\sqrt{2}}\cos\varphi(1+\sin\varphi)^{-1/2} & \,\frac{1}{\sqrt{2}}\cos\varphi(1-\sin\varphi)^{-1/2}\, & 0\\
\frac{1}{\sqrt{2}}(1+\sin\varphi)^{1/2} & \frac{1}{\sqrt{2}}(1-\sin\varphi)^{1/2} & 0\\
0 & 0 & 1
\end{pmatrix}\!,
\end{equation}
where $\cos\varphi>0$ for the principal branch. Therefore,
\begin{alignat}{2}
\tau^{\tr}g\tau & =-\cos\varphi<0, & \qquad\Sigma^{\tr}g\Sigma & =\cos\varphi\, x^{2}+y^{2}>0,\\
\tau^{\tr}f\tau & =-\lambda\cos\varphi<0, & \Sigma^{\tr}f\Sigma & =\lambda\cos\varphi\, x^{2}+\lambda_{2}y^{2}>0,\\
\tau^{\tr}h\tau & =-\lambda^{1/2}<0, & \Sigma^{\tr}h\Sigma & =\lambda^{1/2}x^{2}+\lambda_{2}^{1/2}y^{2}>0.
\end{alignat}

\paragraph{Type III.}

This is the case where the diagonalization is suppressed by the presence
of a common `saddle' tangent plane for the null cones of $f$ and $g$.
Here, $g$ is given in the null coordinates of $t,y$,
\begin{equation}
g=\begin{pmatrix}\, & 0 & \,0\, & 1 & \,\\[-0.2em]
 & 0 & 1 & 0\\[-0.2em]
 & 1 & 0 & 0
\end{pmatrix}\!,\quad f=\begin{pmatrix}\, & 0 & \,0\, & \lambda & \,\\[-0.2em]
 & 0 & \lambda & 1\\[-0.2em]
 & \lambda & 1 & 0
\end{pmatrix}\!,\quad h=\begin{pmatrix}\, & 0 & \,0\, & \lambda^{1/2} & \,\\
 & 0 & \lambda^{1/2} & \,\frac{1}{2}\lambda^{-1/2}\\
 & \lambda^{1/2} & \,\frac{1}{2}\lambda^{-1/2} & \,\frac{-1}{8}\lambda^{-3/2}
\end{pmatrix}\!.
\end{equation}
A congruence $X$ which diagonalizes $h$ reads,
\begin{equation}
X=\lambda^{-1/4}\begin{pmatrix}
\frac{\sqrt{2}}{4}\lambda^{-1} && 0 && \frac{\sqrt{2}}{2}\lambda^{-1}\\[0.2em]
\frac{\sqrt{2}}{3} && 1 && -\frac{\sqrt{2}}{3}\\[0.2em]
-\frac{2\sqrt{2}}{3}\lambda && 0 && \frac{2\sqrt{2}}{3}\lambda
\end{pmatrix}\!.
\end{equation}
This gives,
\begin{alignat}{2}
\tau^{\tr}g\tau & =-{\tfrac{4}{9}}\lambda^{-1/2}<0, & \qquad\Sigma^{\tr}g\Sigma & ={\tfrac{1}{9}}\lambda^{-1/2}\left({(3x-\sqrt{2}y)}^{2}+12y^{2}\right)>0,\\
\tau^{\tr}f\tau & =-{\tfrac{4}{3}}\lambda^{1/2}<0, & \Sigma^{\tr}f\Sigma & ={\tfrac{1}{3}}\lambda^{1/2}\left(2x^{2}+{(x+\sqrt{2}y)}^{2}\right)>0,\\
\tau^{\tr}h\tau & =-1<0, & \Sigma^{\tr}h\Sigma & =x^{2}+y^{2}>0.
\end{alignat}

\paragraph{Type IV.}

This is a special case where,
\begin{equation}\label{eq:suf-t4}
g=\diag(-1,1,1),\quad f=\diag(\lambda,-\lambda,\lambda_{2})\quad 
h=\diag(\lambda^{1/2}\begin{pmatrix}0 & -1\\[-0.2em]
1 & \phantom{-}0
\end{pmatrix}\!,\lambda_{2}^{1/2}).
\end{equation}
The relation $-t^{2}+x^{2}=-(t^{2}-x^{2})$ 
implies that the interior one null cone is in the exterior of the other,
which does not allow the existence of a common timelike vector nor
a common spacelike surface element.
The intersection of the cones is set of null vectors $\tau=(t,\pm t,0)$.
The two common surface elements are null hyperplanes $\Sigma_{1}=(-x,x,y)$
and $\Sigma_{2}=(x,-x,y)$.


\section{Proof of the necessary condition}
\label{app:converse}

In this section, we show that if $\gmn$ and $\fmn$ are causally coupled
or null coupled, then the matrix $g^{-1}f$ has a real square root.
For the null coupled case, the existence of two null hyperplanes
makes possible to write the metrics in the form \eqref{eq:suf-t4},
which is obviously sufficient for the existence of a real square root.

On the other hand, for the causally coupled case, since there exists a common
spacelike hypersurface, we can find a coordinate
frame where we have the simultaneous proper 3+1 decomposition of both
the metrics $g$ and $f$, given by (in matrix notation),
\begin{equation}
  g = \begin{pmatrix}
      -N^{2}+\nu^{\tr}\tilde{g}\nu\, &~& \nu^{\tr}\tilde{g}\\
      \tilde{g}\nu\, && \tilde{g}
    \end{pmatrix}\!,
  \qquad 
  f = \begin{pmatrix}
      -M^{2}+\mu^{\tr}\tilde{f}\mu\, &~& \mu^{\tr}\tilde{f}\\
      \tilde{f}\mu\, && \tilde{f}
    \end{pmatrix}\!.
\end{equation}
Here, $N$ and $M$ are the lapses, $\nu$ and $\mu$ are the shifts,
and $\tilde{g}$ and $\tilde{f}$ are the spatial restrictions of
$g$ and $f$, respectively. Now, let $X=(t,\,x)^{\tr}$ be a nonzero
vector in the tangent space which belongs to the null cone of the
metric $g$ that is defined by $X^{\tr}gX=0$, or equivalently,
\begin{equation}
    (\nu+x/t)^{\tr} \; \tilde{g} \; (\nu+x/t)=N^{2}.\label{eq:nc-1}
\end{equation}
Note that $x$ and $\nu$ are spatial vectors.
Since $\tilde{g}$ is positive definite, we can always find (by Cholesky
decomposition) a lower-triangular $e$ (a spatial vielbein) such that
$\tilde{g}=e^{\tr}\tilde{\delta}e$ where $\tilde{\delta}$ denotes
the spatial Euclidean metric of the local Lorentz frame. In terms
of $e$, (\ref{eq:nc-1}) can be written in a more symmetric form, 
\begin{equation}
   \left(N^{-1}e(\nu+x/t)\right)^{\tr}\,
   \tilde{\delta}\,
   \left(N^{-1}e(\nu+x/t)\right)=\alpha^{2},\label{eq:nc-2}
\end{equation}
where we introduced a real parameter $\alpha$ for which $\alpha^{2}=1$
defines the null cone, while $\alpha^{2}<1$ defines its interior. 
By further introducing a local Lorentz frame spatial vector $u$, 
\begin{equation}
    u \equiv N^{-1}e \, (\nu+x/t),
\end{equation}
we can write (\ref{eq:nc-2}) as $u^{\tr}\tilde{\delta}u=\alpha^{2}$,
and parameterize the null cone as the set of all null rays,
\begin{equation}
    x=(Ne^{-1}u-\nu)\,t,\qquad t\in\mathbb{R}\backslash\{0\},
\end{equation}
that is generated by vectors $u$ on a unit sphere 
$|u|^{2} \equiv u^{\tr}\tilde{\delta}u=u^{a}\delta_{ab}u^{b}=\alpha^{2}=1$.
In such case, the open set (a soft ball) $\alpha^{2}<1$ correspondingly
defines the null cone interior.\footnote{%
  One can also fix $|u|^{2}=1$ and change $N\to\alpha N$.
} The local vector $u$ is, of course, determined up to an $\mathrm{O}(3)$
rotation that keeps the spatial Euclidean metric $\tilde{\delta}$
invariant, i.e., $\tilde{\delta}=R^{\tr}\tilde{\delta}R$. Geometrically,
the spatial metric $\tilde{g}$ (or the corresponding $e$) will deform
a $S^{2}$-sphere $|u|^{2}=1$ (``the light front'' at fixed time)
into an ellipsoid, further scaled by the lapse $N$ (the local passage
of time) which will finally be centered at the shift $-\nu$ (that
corresponds to shear of time relative to the spacelike hypersurface)

Now, let us parameterize the interiors of null cones $g$ and $f$
by $u_{1}$ and $u_{2}$, respectively. Then, if the interiors of
the null cones intersect, there exists some $\alpha$ for which we
can find a common $u=-Ru_{1}=u_{2}$ (up to a spatial rotation $R$) 
so that,
\begin{equation}
    x/t=Ne^{-1}u-\nu=-M(Rm)^{-1}u-\mu,
\end{equation}
where $u^{\tr}\tilde{\delta}u=\alpha^{2}<1$. This yields, 
\begin{equation}
    \nu-\mu=\left(Ne^{-1}+M(Rm)^{-1}\right)u,
\end{equation}
which is the sufficient condition for the existence of the real square
root \cite{HKS}.


\section{Geometric mean of symmetric matrices}
\label{app:geomMean}

Consider a matrix function $F(g^{-1}f)$ for an arbitrary scalar stem
function $F(X)$, and also define,
\begin{equation}
h_{F}\equiv gF(g^{-1}f).
\end{equation}
Accordingly,
\begin{align}
Z^{\tr}h_{F}Z & =\diag(\epsilon_{1}E_{1}F(J_{1}),\dots,
\epsilon_{q}E_{q}F(J_{q}),F(C_{q+1}),\dots,F(C_{p})).\label{eq:cpt-5}
\end{align}
The upper triangular strip form in (\ref{eq:SqrtFunc}) implies
that $h_{F}$ in (\ref{eq:cpt-5}) is symmetric, $h_{F}=h_{F}^{\tr}$,
for any stem function $F$.\footnote{%
  Note that this holds even for nonprimary functions
  (see Section 1.4 in \cite{Higham08}).
}
Hence, a covariant tensor associated with $h_F$ can function as a metric.
When $F$ is on the principal branch, $h_F$ has the same signature as
$f$ and $g$. 

Now, take the specific stem function,
\begin{equation}
F(X) = X^{\xi} \equiv \exp(\xi\log X),
\end{equation}
for a real $\xi$. For this stem, we can define the composite metric,
\begin{equation}
g_{\xi}\equiv g \, (g^{-1}f)^{\xi},
\end{equation}
where, in particular: $g_{0}=g$, $g_{1}=f$, and for $\xi=1/2$ (taking
the principal square root),
\begin{equation}
h\equiv g_{1/2} = g \, (g^{-1}f)^{1/2}
  = f\, (f^{-1}g)^{1/2} = f\op{\#}g=g\op{\#}f.\label{eq:def-h}
\end{equation}
Here, we used $A\op{\#}B$ to denote the \textit{geometric mean} of $A$ and
$B$ (Section 2.4 in \cite{Higham08}).  For the null cones, the term
geometric mean is well justified, since one has,
\begin{equation}
 H^{2}=\gamma^{-1}(M^{2}N^{2})^{1/2},\quad
\det\tilde{h}=\gamma (\det\tilde{f}\,\det\tilde{g})^{1/2},
\end{equation}
in terms of the spatial parts $\tilde{f}$, $\tilde{g}$, $\tilde{h}$
and the lapses $M$, $N$ and $H$ of $f$, $g$ and $h$, respectively,
where the $\gamma$-factor ranges $0<\gamma^{-1}=\sqrt{x}\le1$, 
\cite{HKS}.

It is easy to verify that the set of all $g_{\xi}$ stays closed under 
the binary operation,
\begin{equation}
g_{\xi} = g_{\xi_{1}} \op{\#} g_{\xi_{2}}
= g_{\xi_{1}}(g_{\xi_{1}}^{-1}g_{\xi_{2}})^{1/2}
= g\,(g^{-1}f)^{(\xi_{1}+\xi_{2})/2}
= g_{(\xi_{1}+\xi_{2})/2}\,,
\end{equation}
especially on the segment $\xi,\xi_{1},\xi_{2}\in[0,1]$. 
This makes possible the ordering 
$g_{\xi_1}\prec g_{\xi_2}\Leftrightarrow\xi_1<\xi_2$, and
$g=g_{0}\prec g_{\xi}\prec g_{1}=f$ for $0<\xi<1$.
Finally, note that the null cones of all metrics $g_{\xi}$
are causally coupled to each other.


\section{Absence of relative time orientation flip}
\label{sec:a-Orientability} 

As shown in the main text, Type IV metrics are obtainable as a limit
of Type IIb when $\theta\to\pi$. Although, as argued in the text, Type
IV metrics cannot arise as valid solutions, on the face of it,
crossing $\theta=\pi$ could lead to a problem with time
orientability. Here we show that this is not the case.  Geometrically,
in terms of IIb null cones in Figure \ref{fig:jbs}, $\theta$ is the
angle between the left edge of the blue cone and the right edge of the
red cone. Initially, for $\theta<\pi$, the null cones have common
timelike directions within their intersection. As $\theta \to \pi$,
the intersection shrinks to zero, and the null cones touch along two
null directions (Type IV in Figure \ref{fig:jbs}). One null direction
is a limit of the common timelike vectors while the other null
direction was initially on a spacelike hypersurface. If $\theta$
increases beyond $\pi$ the null cones intersect again, in a new IIb
configuration. However, now what \textit{was} the future cone of $f$
intersects with the past cone of $g$, so seemingly the two metrics
develop opposite time orientations.

This problem will not arise if the transition through $\theta=\pi$ is
accompanied by a time reversal for the $f$ metric. This is not
prohibited for the following reason. The new common time directions
for $\theta>\pi$ are not continuations of the old timelike directions
for $\theta<\pi$. Rather, they arise from the Type IV null direction
that originated from spacelike directions for $\theta<\pi$.
Similarly, the spatial hypersurfaces for $\theta>\pi$ arise from the
Type IV null direction that was initially timelike (In principle, a
2+2 decomposition can describe such a transition through
$\theta=\pi$). This lack of continuity in the common time directions
allows for a reinterpretation of time orientation. Below we show that
specifying a branch for the square root matrix indeed implements this
mechanism, ensuring that the time orientations of $f$ and $g$ remain
compatible.

When a Type IIb configuration crosses $\theta=\pi$ into a new Type IIb
configuration, one encounters a potential problem with the relative
time-orientability of the two metrics, as discussed above. We now show
that the transition through $\theta=\pi$ is accompanied by time and
space reflections of the $f$ metric (in our parametrization) so that
the relative time orientations of the two metrics remain
unchanged. This can be seen directly by noting that while the $f$
metric of Type IIb (as parametrized in \eqref{type2b}) is continuous
at $\theta=\pi$, its vielbein is not and exhibits the time and space
reflections. The $f$ vielbein can be easily obtained as follows.
Associated with a complex number $z$,  consider the matrix,
\begin{equation}
  \mcC{z} \equiv
  \begin{pmatrixr} \Re z &\; -\Im z \\[-0.2em] \Im z &\; \Re z \end{pmatrixr}\!,
\end{equation}
where $\mcC{1}$ is the identity matrix. Then the following
equations hold,
\begin{equation}
  \mcC{z} \mcC{w} = \mcC{z w}, \quad
  {\mcC{z}}^{-1}  = \mcC{z^{-1}}, \quad
  {\mcC{z}}^{\tr} = \mcC{\bar{z}}, \quad
  \sqrt{\mcC{z}}  = \mcC{\sqrt{z}} \;.
\end{equation}
where the last equation follows from \eqref{eq:sqrtIIb}. The
reversal matrix associates a metric to $z$,  
\begin{equation}
\hat f(z) 
\equiv
  \begin{pmatrix} 0 & 1 \\[-0.2em] 1 & 0 \end{pmatrix} \mcC{z}, \quad
 \hat f(z) =\hat f(1) \mcC{z}, \quad 
\hat f(z) = {\hat f}^{\tr} (z)\;.
\end{equation}
Note that  $\hat f(1)\equiv \hat \eta$ is the 2$\times$2 Minkowski
metric in the  null frame and $\hat f(z)$ is the 2$\times$2
block of the metric $f$ in \eqref{type2b}.   Since $\mcC{z} = \mcC{\sqrt{z}}
\mcC{\sqrt{z}}$ and $\hat f(1) \mcC{\sqrt{z}}$ is symmetric, we can
write, 
\begin{equation}
\hat f = {\mcC{\sqrt{z}}}^{\tr} \hat f(1) \mcC{\sqrt{z}} 
\end{equation}
Hence, up to local Lorentz transformations, the vielbein of $\hat f=
\hat L^\tr \hat \eta \hat L$ is given by 
\begin{align}
\hat L = \mcC{\sqrt{z}}\,.    
\label{eq:c-g1}
\end{align}
But  $ \mcC{\sqrt{z}}$ is the square root matrix \eqref{eq:sqrtIIb}
which is discontinuous  at $\theta=\pi$.  Then, from
\eqref{eq:Re-a+ib}  and \eqref{eq:Im-a+ib} it is obvious that,
\begin{align}
\lim_ {~~\theta\to \pi^-}\hat L  =\begin{pmatrixr} 0 &\; -1
  \\[-0.2em] 1 &\; 0 \end{pmatrixr}\,, \qquad
\lim_{~~\theta\to \pi^+}\hat L  = - \begin{pmatrixr} 0 &\; -1 \\[-0.2em]
  1 &\; 0 \end{pmatrixr}.
\end{align}
The sign change across $\theta=\pi$ is due to a reflection in the time
and in a space direction and ensures that the future cones of $f$ and
$g$ continue intersecting. Embedding this setup in four dimensions and
taking the congruence $Z$ into account  replaces the $f$ vielbein by
$LZ$ but does not change the above conclusions.



\clearpage
\begin{thebibliography}{10}

\bibitem{FP}
  M.~Fierz and W.~Pauli,
  Proc.\ Roy.\ Soc.\ Lond.\ A {\bf 173} (1939) 211.
  \doi{doi:10.1098/rspa.1939.0140}

\bibitem{BD}
  D.~G.~Boulware and S.~Deser,
  Phys.\ Rev.\ D {\bf 6} (1972) 3368.
  \doi{doi:10.1103/PhysRevD.6.3368}

\bibitem{Isham:1971gm}
  C.~J.~Isham, A.~Salam and J.~A.~Strathdee,
  Phys.\ Rev.\ D {\bf 3} (1971) 867.
  \doi{doi:10.1103/PhysRevD.3.867}

\bibitem{Zumino} B.~Zumino, 
  ``Effective Lagrangians and Broken Symmetries,'' Lectures on
  Elementary Particles and Quantum Field Theory v.2,
  Brandeis Univ., Cambridge, MA, 1970, pp.\,437--500

\bibitem{Boulanger:2000rq}
  N.~Boulanger, T.~Damour, L.~Gualtieri and M.~Henneaux,
  Nucl.\ Phys.\ B {\bf 597} (2001) 127
  \doi{doi:10.1016/S0550-3213(00)00718-5}
  [hep-th/0007220].

\bibitem{dRGT}
  C.~de Rham, G.~Gabadadze and A.~J.~Tolley,
  Phys.\ Rev.\ Lett.\  {\bf 106} (2011) 231101
  \doi{doi:10.1103/PhysRevLett.106.231101}
  [arXiv:1011.1232 [hep-th]].

\bibitem{HR1106}
  S.~F.~Hassan and R.~A.~Rosen,
  Phys.\ Rev.\ Lett.\  {\bf 108} (2012) 041101
  \doi{doi:10.1103/PhysRevLett.108.041101}
  [arXiv:1106.3344 [hep-th]].

\bibitem{HR1109}
  S.~F.~Hassan and R.~A.~Rosen,
  JHEP {\bf 1202} (2012) 126
  \doi{doi:10.1007/JHEP02(2012)126}
  [arXiv:1109.3515 [hep-th]].

\bibitem{HR1111}
  S.~F.~Hassan and R.~A.~Rosen,
  JHEP {\bf 1204} (2012) 123
  \doi{doi:10.1007/JHEP04(2012)123}
  [arXiv:1111.2070 [hep-th]].

\bibitem{ArkaniHamed:2002sp}
  N.~Arkani-Hamed, H.~Georgi and M.~D.~Schwartz,
  Annals Phys.\  {\bf 305} (2003) 96
  \doi{doi:10.1016/S0003-4916(03)00068-X}
  [hep-th/0210184].

\bibitem{Creminelli:2005qk}
  P.~Creminelli, A.~Nicolis, M.~Papucci and E.~Trincherini,
  JHEP {\bf 0509} (2005) 003
  \doi{doi:10.1088/1126-6708/2005/09/003}
  [hep-th/0505147].

\bibitem{Ostrogradsky}
  M.~Ostrogradsky,
  Mem. Ac. St. Petersbourg {\bf VI 4}, 385 (1850)

\bibitem{dRG}
  C.~de Rham and G.~Gabadadze,
  Phys.\ Rev.\ D {\bf 82} (2010) 044020
  \doi{doi:10.1103/PhysRevD.82.044020}
  [arXiv:1007.0443 [hep-th]].

\bibitem{HR1103}
  S.~F.~Hassan and R.~A.~Rosen,
  JHEP {\bf 1107} (2011) 009
  \doi{doi:10.1007/JHEP07(2011)009}
  [arXiv:1103.6055 [hep-th]].

\bibitem{HRS1109}
  S.~F.~Hassan, R.~A.~Rosen and A.~Schmidt-May,
  JHEP {\bf 1202} (2012) 026
  \doi{doi:10.1007/JHEP02(2012)026}
  [arXiv:1109.3230 [hep-th]].

\bibitem{HSvS1203}
  S.~F.~Hassan, A.~Schmidt-May and M.~von Strauss,
  Phys.\ Lett.\ B {\bf 715} (2012) 335
  \doi{doi:10.1016/j.physletb.2012.07.018}
  [arXiv:1203.5283 [hep-th]].

\bibitem{Comelli:2012vz}
  D.~Comelli, M.~Crisostomi, F.~Nesti and L.~Pilo,
  Phys.\ Rev.\ D {\bf 86} (2012) 101502
  \doi{doi:10.1103/PhysRevD.86.101502}
  [arXiv:1204.1027 [hep-th]].

\bibitem{Kluson}
  J.~Kluson,
  Phys.\ Rev.\ D {\bf 86} (2012) 044024
  \doi{doi:10.1103/PhysRevD.86.044024}
  [arXiv:1204.2957 [hep-th]].

\bibitem{cedrik}
   C.~Deffayet, J.~Mourad and G.~Zahariade,
  JCAP {\bf 1301} (2013) 032
  \doi{doi:10.1088/1475-7516/2013/01/032}
  [arXiv:1207.6338 [hep-th]].

\bibitem{Kugo:2014hja}
  T.~Kugo and N.~Ohta,
  PTEP {\bf 2014} (2014) 043B04
  \doi{doi:10.1093/ptep/ptu046}
  [arXiv:1401.3873 [hep-th]].

\bibitem{HSvS1515}
  S.~F.~Hassan, A.~Schmidt-May and M.~von Strauss,
  JHEP {\bf 1305} (2013) 086
  \doi{doi:10.1007/JHEP05(2013)086}
  [arXiv:1208.1515 [hep-th]].

\bibitem{deRrev}
  C.~de Rham,
  Living Rev.\ Rel.\  {\bf 17} (2014) 7
  \doi{doi:10.12942/lrr-2014-7}
  [arXiv:1401.4173 [hep-th]].

\bibitem{SvSrev}
  A.~Schmidt-May and M.~von Strauss,
  J.\ Phys.\ A {\bf 49} (2016) no.18,  183001
  \doi{doi:10.1088/1751-8113/49/18/183001}
  [arXiv:1512.00021 [hep-th]].

\bibitem{Visser}
  P.~Mart\'{i}n-Moruno, V.~Baccetti and M.~Visser,
  \doi{doi:10.1142/9789814623995\_0144}
  arXiv:1302.2687.

\bibitem{Akrami:2015qga}
  Y.~Akrami, S.~F.~Hassan, F.~K\"onnig, A.~Schmidt-May and A.~R.~Solomon,
  Phys.\ Lett.\ B {\bf 748} (2015) 37
  \doi{doi:10.1016/j.physletb.2015.06.062}
  [arXiv:1503.07521 [gr-qc]].

\bibitem{HSvS1507}
  S.~F.~Hassan, A.~Schmidt-May and M.~von Strauss,
  Class.\ Quant.\ Grav.\  {\bf 33} (2016) no.1,  015011
  \doi{doi:10.1088/0264-9381/33/1/015011}
  [arXiv:1507.06540 [hep-th]].

\bibitem{Gourgoulhon12}
  E.~Gourgoulhon,
  ``3+1 Formalism in General Relativity.''
  Springer (2012)
  \doi{doi:10.1007/978-3-642-24525-1}
  
\bibitem{Choque-Bruhat}
  Y.~Choquet-Bruhat and R.~P.~Geroch,
  Commun.\ Math.\ Phys.\  {\bf 14} (1969) 329.
  \doi{doi:10.1007/BF01645389}
\bibitem{Bernal:2003jb}
  A.~N.~Bernal and M.~Sanchez,
  Commun.\ Math.\ Phys.\  {\bf 243} (2003) 461
  \doi{doi:10.1007/s00220-003-0982-6}
  [gr-qc/0306108].
\bibitem{Dirac}
  P.~A.~M.~Dirac,
  Proc.\ Roy.\ Soc.\ Lond.\ A {\bf 246} (1958) 333.
  \doi{doi:10.1098/rspa.1958.0142}

\bibitem{ADM60}
  R.~L.~Arnowitt, S.~Deser and C.~W.~Misner,
  Phys.\ Rev.\  {\bf 117} (1960) 1595.
  \doi{doi:10.1103/PhysRev.117.1595}

\bibitem{Higham08}
  N.~J.~Higham,
  ``Functions of Matrices: Theory and Computation.''
  SIAM, 2008.
  \doi{doi:10.1137/1.9780898717778}

\bibitem{Higham87}
  N.~J.~Higham,
  ``Computing real square roots of a real matrix.''
  Linear Algebra Appl. 88-89 (1987), 405--430.
  \doi{doi:10.1016/0024-3795(87)90118-2 }

\bibitem{Horn94}
  R.~A.~Horn and C.~R.~Johnson. 
  ``Topics in Matrix Analysis.''
  Cambridge Univ. Pr. (1994)
  \doi{doi:10.1017/CBO9780511840371}

\bibitem{Uhlig73}
  F.~Uhlig,
  ``Simultaneous block diagonalization of two real symmetric matrices.''
  Linear Algebra Appl. 7 (1973) 4, 281--289.
  \doi{doi:10.1016/S0024-3795(73)80001-1}

\bibitem{Uhlig76}
  F.~Uhlig,
  ``A canonical form for a pair of
  real symmetric matrices that generate a nonsingular pencil.''
  Linear Algebra Appl. 14 (1976) 3, 189--209.
  \doi{doi:10.1016/0024-3795(76)90066-5}

\bibitem{Uhlig79}
  F.~Uhlig,
  ``A recurring theorem about pairs of quadratic forms 
  and extensions: a survey.''
  Linear Algebra Appl. 25 (1979) 0, 219--237.
  \doi{doi:10.1016/0024-3795(79)90020-X}

\bibitem{Gant59v2}
  F.~R.~Gantmacher,
  ``The Theory of Matrices,'' Vol. 2.
  Chelsea (1959)

\bibitem{Hong86}
  Y.~P.~Hong, R.~A.~Horn, and C.~R.~Johnson,
  ``On the reduction of pairs of Hermitian or symmetric matrices
  to diagonal form by congruence.'' 
  Linear Algebra Appl. 73 (1986), 213--226.
  \doi{doi:10.1016/0024-3795(86)90241-7}

\bibitem{Horn90}
  R.~A.~Horn and C.~R.~Johnson,
  ``Matrix Analysis.''
  Cambridge Univ. Pr. (1990)
  \doi{doi:10.1017/CBO9781139020411}

\bibitem{Baccetti:2012ge}
  V.~Baccetti, P.~Martin-Moruno and M.~Visser,
  JHEP {\bf 1208} (2012) 108
  \doi{doi:10.1007/JHEP08(2012)108}
  [arXiv:1206.4720 [gr-qc]].

\bibitem{DSS1}
  R.~A.~d'Inverno and J.~Stachel, 
  J.\ Math.\ Phys. {\bf 19} (1978) 2447.
  \doi{doi:10.1063/1.523650}

\bibitem{DSS2}
  R.~A.~d'Inverno and J.~Smallwood,
  Phys.\ Rev.\ D {\bf 22} (1980) 1233.
  \doi{doi:10.1103/PhysRevD.22.1233}
  
\bibitem{HSvS1212}
  S.~F.~Hassan, A.~Schmidt-May and M.~von Strauss,
  Class.\ Quant.\ Grav.\  {\bf 30} (2013) 184010
  \doi{doi:10.1088/0264-9381/30/18/184010}
  [arXiv:1212.4525 [hep-th]].
 
\bibitem{HKS}
  S.~F.~Hassan, M.~Kocic and A.~Schmidt-May,
  \href{https://arxiv.org/abs/1409.1909}{[arXiv:1409.1909 [hep-th]]}.

\bibitem{Bernard:2015mkk}
  L.~Bernard, C.~Deffayet and M.~von Strauss,
  JCAP {\bf 1506} (2015) 038
  \doi{doi:10.1088/1475-7516/2015/06/038}
  [arXiv:1504.04382 [hep-th]].

\bibitem{IO}
  K.~Izumi and Y.~C.~Ong,
  Class.\ Quant.\ Grav.\  {\bf 30} (2013) 184008
  \doi{doi:10.1088/0264-9381/30/18/184008}
  [arXiv:1304.0211 [hep-th]].

\bibitem{DIOWredux}
  S.~Deser, K.~Izumi, Y.~C.~Ong and A.~Waldron,
  Phys.\ Lett.\ B {\bf 726} (2013) 544
  \doi{doi:10.1016/j.physletb.2013.09.001}
  [arXiv:1306.5457 [hep-th]].

\bibitem{DIOW1312}
  S.~Deser, K.~Izumi, Y.~C.~Ong and A.~Waldron,
  \doi{doi:10.1142/9789814590112\_0029}
  arXiv:1312.1115 [hep-th].

\bibitem{DSWZ1408}
  S.~Deser, M.~Sandora, A.~Waldron and G.~Zahariade,
  Phys.\ Rev.\ D {\bf 90} (2014) no.10,  104043
  \doi{doi:10.1103/PhysRevD.90.104043}
  [arXiv:1408.0561 [hep-th]].

\bibitem{DIOWproblems}
  S.~Deser, K.~Izumi, Y.~C.~Ong and A.~Waldron,
  Mod.\ Phys.\ Lett.\ A {\bf 30} (2015) 1540006
  \doi{doi:10.1142/S0217732315400064}
  [arXiv:1410.2289 [hep-th]].

\bibitem{Geroch}
  R.~Geroch,
  ``Faster Than Light?,''
  \href{https://arxiv.org/abs/1005.1614}{[arXiv:1005.1614 [gr-qc]]}.

\bibitem{Babichev:2007dw}
  E.~Babichev, V.~Mukhanov and A.~Vikman,
  JHEP {\bf 0802} (2008) 101
  \doi{doi:10.1088/1126-6708/2008/02/101}
  [arXiv:0708.0561 [hep-th]].

\bibitem{Schuller:2016onj}
  F.~P.~Schuller, N.~Stritzelberger, F.~Wolz and M.~D\"{u}ll,
  ``Gravitational closure of matter field equations,''
  \href{https://arxiv.org/abs/1611.08878}{[arXiv:1611.08878 [gr-qc]]}.

\bibitem{Drummond:2013ida}
  I.~T.~Drummond,
  Phys.\ Rev.\ D {\bf 88} (2013) no.2,  025009
  \doi{doi:10.1103/PhysRevD.88.025009}
  [arXiv:1303.3126 [hep-th]].

\bibitem{Volkov1}
  M.~S.~Volkov,
  Phys.\ Rev.\ D {\bf 90} (2014) no.2,  024028
  \doi{doi:10.1103/PhysRevD.90.024028}
  [arXiv:1402.2953 [hep-th]].

\bibitem{Volkov2}
  M.~S.~Volkov,
  Phys.\ Rev.\ D {\bf 90} (2014) no.12,  124090
  \doi{doi:10.1103/PhysRevD.90.124090}
  [arXiv:1404.2291 [hep-th]].

\bibitem{HSvS1407}
  S.~F.~Hassan, A.~Schmidt-May and M.~von Strauss,
  Int.\ J.\ Mod.\ Phys.\ D {\bf 23} (2014) no.13,  1443002
  \doi{doi:10.1142/S0218271814430020}
  [arXiv:1407.2772 [hep-th]].

\bibitem{HR}
  K.~Hinterbichler and R.~A.~Rosen,
  JHEP {\bf 1207} (2012) 047
  \doi{doi:10.1007/JHEP07(2012)047}
  [arXiv:1203.5783 [hep-th]].
 
\bibitem{Bojowald}
  M.~Bojowald,
  ``Canonical Gravity and Applications.''
  Cambridge Univ. Pr. (2010)
  \doi{doi:10.1017/CBO9780511921759}

\bibitem{Camanho:2016opx}
  X.~O.~Camanho, G.~Lucena Gomez and R.~Rahman,
  \href{https://arxiv.org/abs/1610.02033}{[arXiv:1610.02033 [hep-th]]}.
 
\bibitem{Alexandrov}
  S.~Alexandrov,
  Gen.\ Rel.\ Grav.\  {\bf 46} (2014) 1639
  \doi{doi:10.1007/s10714-013-1639-1}
  [arXiv:1308.6586 [hep-th]].

\bibitem{HL}
  S.~F.~Hassan and A.~Lundkvist,
  \href{https://arxiv.org/abs/1802.07267 [hep-th]}{[arXiv:1802.07267 [hep-th]]}.

\bibitem{Bernard:2015uic}
  L.~Bernard, C.~Deffayet, A.~Schmidt-May and M.~von Strauss,
  Phys.\ Rev.\ D {\bf 93} (2016) no.8,  084020
  \doi{doi:10.1103/PhysRevD.93.084020}
  [arXiv:1512.03620 [hep-th]].

\bibitem{Deffayet:2012zc}
  C.~Deffayet, J.~Mourad and G.~Zahariade,
  JHEP {\bf 1303} (2013) 086
  \doi{doi:10.1007/JHEP03(2013)086}
  [arXiv:1208.4493 [gr-qc]].

\bibitem{HSvS1204}
  S.~F.~Hassan, A.~Schmidt-May and M.~von Strauss,
  \href{https://arxiv.org/abs/1204.5202}{[arXiv:1204.5202 [hep-th]]}.
  
\bibitem{AH1609}
  L.~Apolo and S.~F.~Hassan,
  Class.\ Quant.\ Grav.\  {\bf 34} (2017) no.10,  105005
  \doi{doi:10.1088/1361-6382/aa69f7}
  [arXiv:1609.09514 [hep-th]].

\bibitem{Cheung:2016yqr}
  C.~Cheung and G.~N.~Remmen,
  JHEP {\bf 1604} (2016) 002
  \doi{doi:10.1007/JHEP04(2016)002}
  [arXiv:1601.04068 [hep-th]].

\bibitem{Comelli:2015ksa}
  D.~Comelli, M.~Crisostomi, K.~Koyama, L.~Pilo and G.~Tasinato,
  Phys.\ Rev.\ D {\bf 91} (2015) no.12,  121502
  \doi{doi:10.1103/PhysRevD.91.121502}
  [arXiv:1505.00632 [hep-th]].

\end{thebibliography}
\end{document}